\PassOptionsToPackage{pdfpagelabels=false}{hyperref}

\documentclass[twocolumn]{aastex631}

\usepackage{amsmath}
\usepackage{amssymb}
\usepackage{xcolor}

\begin{document}
\title{}

\title{Simulated 3D $^{56}$Ni Distributions of Type IIp Supernovae} 


\author[0000-0003-1938-9282]{David Vartanyan}
\affiliation{Department of Physics, University of Idaho, ID 83843, USA}\affiliation{Carnegie Observatories, 813 Santa Barbara St., Pasadena, CA 91101, USA; NASA Hubble Fellow}
\author[0000-0002-3099-5024]{Adam Burrows}
\affiliation{Department of Astrophysical Sciences, Princeton University, NJ 08544, USA}
\author[0009-0009-3068-9527]{Lizzy Teryoshin}
\affiliation{University of California, San Diego, CA, 92037, USA}
\author[0000-0002-0042-9873]{Tianshu Wang}
\affiliation{ Department of Physics, University of California Berkeley, Berkeley, CA, 94720, USA}
\author[0000-0002-5981-1022]{Daniel Kasen}
\affiliation{Department of Astronomy and Theoretical Astrophysics Center, University of California, Berkeley, CA 94720, USA}\affiliation{Nuclear Science Division, Lawrence Berkeley National Laboratory, Berkeley, CA, 94720, USA}
\author[0000-0002-6543-2993]{Benny T.-H. Tsang}
\affiliation{Department of Astronomy and Theoretical Astrophysics Center, University of California, Berkeley, CA 94720, USA}
\author[0000-0001-5939-5957c]{Matthew S.~B. Coleman}
\affiliation{Research Computing, Princeton University, Princeton, NJ 08544, USA}



   

\begin{abstract}

We present the first three-dimensional study of the asymptotic ejecta distributions for a suite of theoretical Type IIp supernovae originating from red supergiant progenitors. 
We simulate using the radiation-hydrodynamic code F{\sc{ornax}} from core bounce through the first seconds of the neutrino-driven explosion and then follow using a hydrodynamic variant of the code FLASH until shock breakout of the star and through to homologous expansion of the ejecta into the circumstellar environment. Our studied progenitor models range from 9 to 25 M$_{\odot}$, with explosion energies spanning $\sim$0.1$-$1 Bethe. The shock breakout times span the range $\sim$1$-$4 days, with a breakout time spread by direction ranging from hours to over a day. We find that the dipole orientation of the $^{56}$Ni ejecta is well-preserved from the first seconds out to shock breakout. The $^{56}$Ni ejecta penetrates through the initially outer oxygen shell, and its global structure is imprinted with small-scale clumping as the ejecta evolve through the stellar envelope. For the majority of our models, the neutron star kick is anti-aligned with the $^{56}$Ni ejecta. Models with strongly dipolar ejecta morphology and a massive hydrogen/helium envelope with an inner boundary located deep see as much as $\sim$70\% of the $^{56}$Ni ejecta mixed into that outer envelope, reaching asymptotic velocities ranging from $\sim$350 to 3200 km s$^{-1}$. Supernovae arising from red supergiant progenitors and exhibiting prominent nickel features generally display significant $^{56}$Ni mixing into the stellar envelope.
\end{abstract}


    

\section{Introduction}

Core-collapse supernovae (CCSNe) illustrate multi-scale asymmetries, from neutrino-driven turbulence at the onset of explosion to large-scale interface and reverse-shock instabilities in the expanding ejecta \citep{muller_janka_pert, 2015ApJ...808..164M,2020ApJ...890...94Y,gabler,sandoval,wongwathanarat2017,burrows2024, 2018MNRAS.479.3675M,vartanyan2025, 2020ApJ...888..111O}. A litany of recent observations highlights both the intrinsic and extrinsic asymmetry in explosion morphology among well-studied supernovae and their remnants \citep{2024ApJ...965L..27M,2024ApJ...968L..18T,2025arXiv250803395L, 2023ApJ...958...95J,2024MNRAS.532.3625M}. 

The origin, evolution, and morphology of metal mixing into the envelope of red supergiants provide insights into the various avenues by which asymmetry is introduced into CCSNe. Filamentary structures in supernova remnant Cassiopeia A (Cas A) can be reproduced with the summed effect of neutrino-heated plumes, Rayleigh-Taylor-like interface and reverse-shock instabilities, and the Ni-bubble effect \citep{orlando2025}. A crucial agency of metal enrichment in the outer CCSN ejecta is the explosion itself that powers differential mixing of the core ejecta into the envelope. Furthermore, three-dimensional studies of the innermost star in the minutes before core collapse \citep{muller2017,Fields2021, vartanyan2022} suggest that pre-collapse perturbations can both facilitate explosion and shape its early structure. Curiously, the development of three-dimensional models for stellar progenitor envelopes \citep{goldberg} can further complicate the explosion narrative from pre-collapse to observation. 

Each of these features $-$ pre-collapse core asymmetries, neutrino-driven turbulence, and structure in the stellar envelope and in the circumstellar medium (CSM) $-$ imprint their signatures onto the final morphological structure of the CCSN.
Asymmetric structure is not simply an added detail in the study of supernovae, but a central aspect of observed and theoretical SN and SNR morphologies. Moreover, prior work connected the structures in observed SNRs with those of the progenitor supernovae and their host environments \citep{orlando-cent}, showing early agreement between models and observations in the CCSNe $^{56}$Ni distribution \citep{2017ApJ...841..127M,2017IAUS..331..148J}.

An important recent finding is that destratification in N49B \citep{2025ApJ...984..185S} and inhomogeneities in Cas A \citep{2025arXiv250707563S} may require pre-collapse burning shell mergers and mixing to yield the high Mg/Ne ratios observed in the ejecta. Observations of excess magnesium in SNR N49B suggest possible pre-collapse mixing of metals into the inner stellar envelope. Similarly, observations in Cas A suggest C/O/Ne shell mergers in the final hours before core collapse, leading to Ne downflows and Si upflows. Famously, the Bochum event in SN1987A $-$ a red bump in the H$\alpha$ emission line near $\sim$4500 km s$^{-1}$ $-$ is best explained by a $^{56}$Ni bullet mixed out into the outer envelope from its production site \citep{1988MNRAS.234P..41H, 2002ApJ...579..671W,1995A&A...295..129U}.

The CSM can also impart structure. The holes and rings in the prominent `Green Monster' in Cas A may be explained by the interaction of the ejecta with shocked circumstellar material \citep{2025A&A...696A.188O, 2024ApJ...965L..27M}, though the precise origin of the latter remains unresolved. Ring- and crown-like structures emerge naturally through the interaction of the ejecta with the reverse shock through hydrodynamic instabilities \citep{orlando-mil,vartanyan2025}.
    
We note that the detection of shock breakout in UV, X-ray, and even $\gamma$-rays \citep{2022ApJ...931...15B}, with instruments like ULTRASAT \citep{ultrasat}, UVEX \citep{uvex}, and the Einstein Probe \citep{ep} in the first hours to days after core collapse will provide unprecedented insight into the collective effect of explosion and mixing asymmetries (e.g., \citealt{2025SCPMA..6839501Y}). 
Additionally, 
improvements in collecting area, efficiency, resolution, and speed might enable the interferometric study of line emission from distant supernovae using a variant of the expanding ejecta method \citep{2025arXiv250420132C,2024arXiv240812462O}.






While we here focus on red supergiant progenitors, the most detailed images of CCSNe and remnants have been of yellow and blue supergiants (Cas A, SN1987A, respectively). Envelope stripping, including through binary evolution, can affect progenitor structure \citep{2021A&A...656A..58L,tauris:15,woosley2019}, explosion outcome \citep{2021ApJ...916L...5V,muller_lowmass}, and breakout signatures \citep{2018MNRAS.479.3675M, wongwathanarat2017,kifonidis2000}. In this regard, more recent studies are providing synthetic light curves and spectra for stripped-envelope supernovae \citep{2025ApJ...979..148L,2024arXiv241020829M,2024MNRAS.527.2185M} to aid in the accurate inference of $^{56}$Ni mass and mixing, ejecta mass, and ejecta velocity in this context \citep{2025arXiv250710648K}. We reserve the study of these supernova types to future work.

This paper is organized as follows. In \S\,\ref{sec:methods}, we summarize our methods and introduce the formalism we use in \S\,\ref{sec:formalism}. We discuss our results on the structure, distribution, and velocity of the mixed ejecta in simulated 3D red supergiant explosions in \S\,\ref{sec:results} and discuss our conclusions in \S\,\ref{sec:conc}.

\section{Methods}\label{sec:methods}
We use the framework established in \cite{vartanyan2025} to evolve in three spatial dimensions seven red supergiant (RSG) models from core collapse using the state-of-the-art supernova code F{\sc{ornax}} \citep{skinner2019} through and post shock-breakout using our variant of FLASH. The neutrino-driven explosions are simulated for $\sim$2$-$7 seconds after core bounce, (depending on the progenitor model) \citep{burrows2024}, after which they are passed to FLASH and followed for days thereafter. The seven progenitors are 9 (9b in our catalog), 11-, 15.01-, 20- 23-, 25-M$_{\odot}$ progenitors, with the first two from \cite{swbj16} and the remaining from \cite{sukhbold2018}. In the FLASH continuation, we use the Helmholtz \citep{2000ApJS..126..501T} equation of state and include self-gravity, with sub-degree angular resolution and radial resolution $\Delta{r}/r$ $\sim$(6$-$7)$\times10^{-3}$. The inner boundary condition of the FLASH simulation incorporates 
the gravity of the residual point mass.  We continue our simulations after the shock breaks out from the red supergiant into a low-density circumstellar material. We add $\sim$0.01 M$_{\odot}$ of CSM out to several times the size of the star, R$_\ast$, following an inverse-square power-law density profile (resulting from a constant spherical wind rate with uniform velocity, see \citealt{2017MNRAS.469L.108M} but also \citealt{2024OJAp....7E..47F,2023A&A...677A.105D}) to study the asymptotic behavior of the ejecta as they approach homologous expansion. Table\,\ref{tab:sn_props} summarizes our models and some salient results.

\section{Formalism}\label{sec:formalism}

We summarize below some of the metrics we will use to characterize the formation and evolution of structures in the explosion ejecta. In this paper, we focus particularly on the distribution of $^{56}$Ni. 

\subsection{Isosurface}
Throughout, we will reference calculations done on an isosurface ($^{56}$Ni or $^{16}$O). For each direction, we locate the most radially extended cell in which the element’s mass fraction exceeds a chosen threshold (e.g., 1$\%$ or 30$\%$, respectively, for $^{56}$Ni or oxygen) to define the isosurface.

\subsection{Core Compactness}

The compactness parameter characterizes the inner core structure and is defined as \citep{2011ApJ...730...70O}:

\begin{equation}
\xi_M= \frac{M/M_{\odot}}{(R(M)/1000\, \mathrm{km})}\, ,
\end{equation}
where the subscript $M$ denotes the mass where the compactness parameter is evaluated. We evaluate the compactness parameter $\xi_{1.75}$ at $M$ = 1.75 M$_{\odot}$. The core compactness is a strong indicator of the neutrino luminosity and the accretion rate, which jointly inform the critical curve for determining explosion outcome. However, core compactness is not a monotonic predictor of explodability \citep{tsang2022,wang}.

\subsection{Spherical Harmonics}

We use the approach outlined in \cite{burrows2012} to
decompose a surface A$(\theta,\phi)$ (such as a $^{56}$Ni isosurface) into spherical harmonic components with coefficients:
\begin{equation}\label{eq:alm}
a_{lm} = \frac{(-1)^{|m|}}{\sqrt{4\pi(2l+1)}} \oint A(\theta,\phi) Y_l^m(\theta,\phi) d\Omega\, ,
\end{equation} 

\subsection{Angular Self-Correlation}
\label{self}

In particular, we are also interested in studying the evolution of the $^{56}$Ni isosurface as a function of time. We follow the formalism of \cite{kuroda2017} and \cite{vartanyan2019} to calculate the time-dependent angle-integrated self-correlation $X\,(t)$ for a surface A$_1(t,\Omega)$

\begin{equation}
X\,(t) =  \frac{\int A_1\,(t_0,\Omega)\, A_1\,(t,\Omega)\, d\Omega}{\sqrt{\int A_1\,(t_0,\Omega)^2 d\Omega  \int A_1\,(t,\Omega)^2\,d\Omega}}\,,
\end{equation}
where t$_0$ is the duration of the F{\sc{ornax}} simulation and the time of mapping to FLASH. This metric ranges between $-$1 and 1, spanning strong anti-correlation, weak correlation, and strong correlation.

\subsection{Mixing}
As in \cite{vartanyan2025}, we define mixing into the hydrogen envelope by calculating the fraction of an isotope's total mass mixed beyond the angle-averaged H/He interface, defined where the mass fractions of the two elements are equal in the gaseous RSG envelope. As noted, we focus on $^{56}$Ni mixing. 

\subsection{Clumping}\label{sec:clumping}
As a function of time, we look at the ratio of the $^{56}$Ni isosurface area to the isotropic equivalent area, defined as the area of $^{56}$Ni were it to be distributed in spherically symmetry at the mean radius of the $^{56}$Ni ejecta, to characterize the clumping of the ejecta due to interface and reverse-shock instabilities.\footnote{We use a Marching cubes algorithm to visualize isosurfaces and calculate the area-weighted mean radius of the $^{56}$Ni ejecta in a polygonal decomposition.} Growth of this normalized $^{56}$Ni surface area to values larger than one indicates the formation of clumps. For models with strong dipolar symmetry, the ratio can also be (briefly) less than one, as the ejecta is funneled into one direction, before the onset of small-scale clumping and mixing through hydrodynamic instabilities.

\section{Results}\label{sec:results}

The explosions from our study of 9$-$25 M$_{\odot}$ progenitors produce $^{56}$Ni yields ranging from \textless0.01$-$0.17 M$_{\odot}$, roughly monotonic with progenitor mass, less roughly with compactness. The breakout times are strongly non-monotonic with mass and range from less than one to four days (see Table\,\ref{tab:sn_props}).
Both our distribution and range of breakout times with progenitor mass are similar to  the results in \cite{2022ApJ...934...67B} (see their Figure 6), with two notable exceptions. First, we see in our 3D simulations a breakout duration by viewing angle ranging from hours to days due to the asymmetry of the shock front. Second, we note that the 11- and 17-M$_{\odot}$ models, which are both highly asymmetric and power high ejecta velocities, experience a much earlier breakout than predicted by \cite{2022ApJ...934...67B}. An additional delay of hours to breakout time may be imparted by large convective fluctuations in the envelope density \citep{goldberg2} (not included here) which may in the case of Type IIb SN progenitors result in luminosity variations spanning ten days \citep{goldberg2025}. 

We find that our exploding models fall into two general categories: those with a compactness $\xi_{1.75}$ below $\sim$0.3, and those with $\xi_{1.75}$ above $\sim$0.7. Our model list is certainly not exhaustive, but generally the CCSN simulation community has found that the stellar models that are most difficult to explode are those of intermediate compactness \citep{burrows2024,vartanyan2024,tsang2022,wang, summa2016,oconnor_couch2018a,oconnor_couch2018b}, or marginally explosive when adding magnetic fields \citep{2025MNRAS.536..280N}.

We do not see a clear trend for lower-mass progenitors experiencing earlier shock breakout. Rather, while lower mass models experience higher shock velocities in the first seconds due to lower ambient ram pressures in the cores from weaker accretion, they also have lower explosion energies. In fact, models with higher core compactness have significantly more neutrino heating, growing faster than linear with compactness \citep{2025MNRAS.540.3885B}. The explosion energy generally grows faster than the progenitor mass, and thus the asymptotic velocity $v^2\propto E/M$ is lower for lower mass progenitors.

We see an inversion of velocities at shock breakout by mass: lower mass models, with an initial higher shock velocity, end up with a lower velocity as the ejecta approach homologous expansion. For instance, the 9-M$_{\odot}$ model has the highest shock velocity ($\sim$15,000 km s$^{-1}$) in the first seconds , and the 17-M$_{\odot}$ model the lowest ($\sim$3000 km$^{-1}$), but at breakout the former has the smallest shock velocity ($\sim$3000 km s$^{-1}$) and the latter the highest ($\sim$7000 km s$^{-1}$). Assuming the explosion energy is largely funneled into kinetic energy at shock breakout (though this is not necessarily the case), the time to breakout scales with the RSG radius, R$_{RSG}$, the shock velocity, $v_s$, and in turn with the explosion energy $E$ and the ejecta mass $M_{ej}$ as  $R_{RSG}/v_s\sim R_{RSG}/(E/M_{ej})^{1/2}$. 

\subsection{Ejecta Morphology}
\begin{figure*}
\centering
    \includegraphics[width=0.49\textwidth]{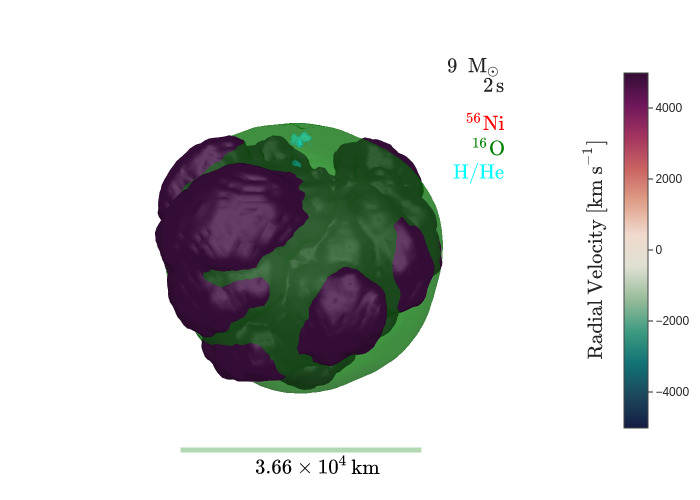}
    \includegraphics[width=0.49\textwidth]{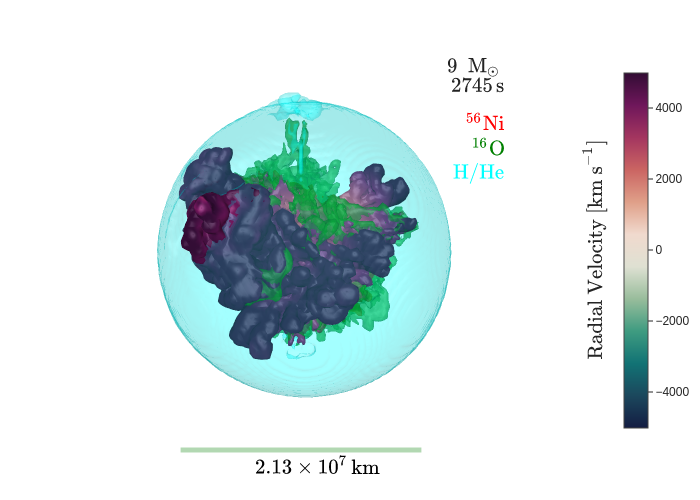}
    \includegraphics[width=0.49\textwidth]{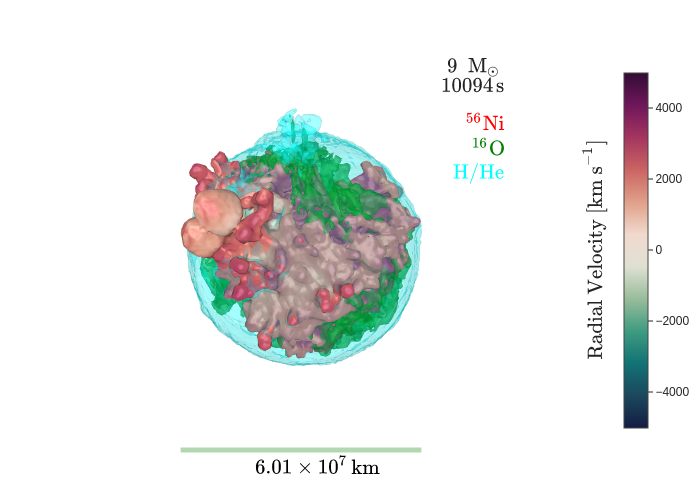}
     \includegraphics[width=0.49\textwidth]{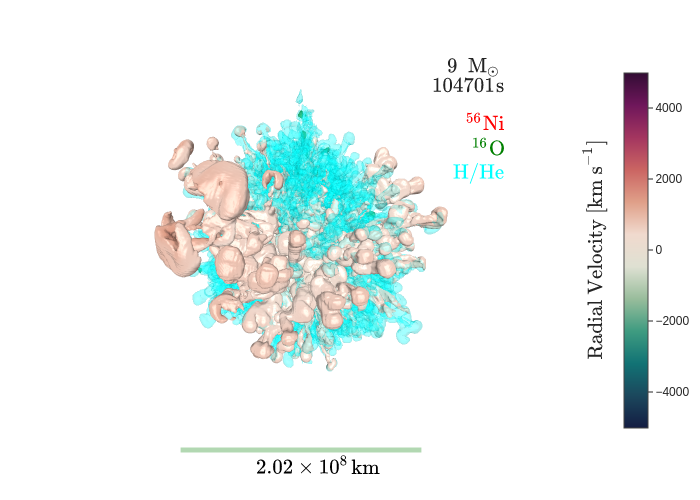}
    \includegraphics[width=0.49\textwidth]{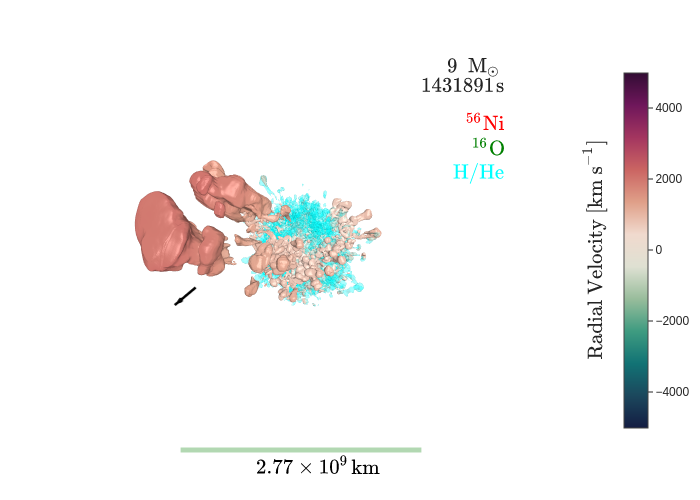} 
   \caption{We illustrate the ejecta morphology from the first seconds of neutrino-driven explosion out to, and beyond, shock breakout from the RSG surface for the 9-M$_{\odot}$ progenitor, showing the 1$\%$ $^{56}$Ni isosurface (colored by radial velocity), 30$\%$ $^{16}$O isosurface (green), and the H/He interface (cyan veil).
   Note the mushroom caps emerging from the reverse shock that forms at the H/He interface, into the H/He interface, and the flattening of the $^{56}$Ni ejecta interior and clumping external. We show the remnant kick velocity direction with a black arrow. The C/O interface is encountered early on for this model. We note how off-center the explosion is, and the large-scale, late-time $^{56}$Ni bullets. The last figure shows its expansion into a low-density CSM.}
    \label{fig:9ni56}
\end{figure*}

\begin{figure*}
\centering
    \includegraphics[width=0.49\textwidth]{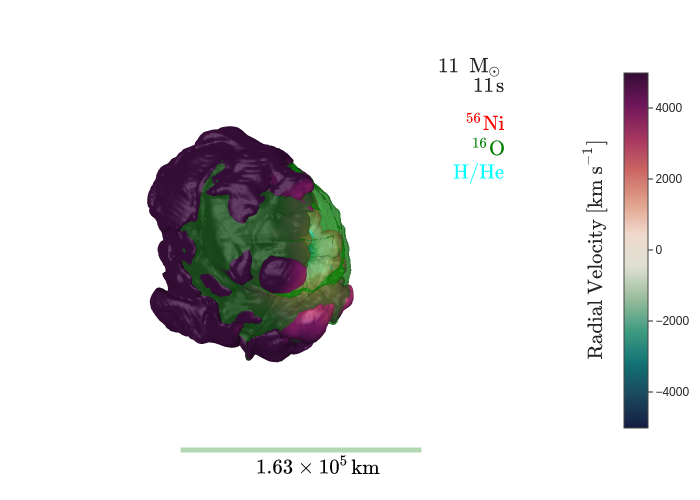}
    \includegraphics[width=0.49\textwidth]{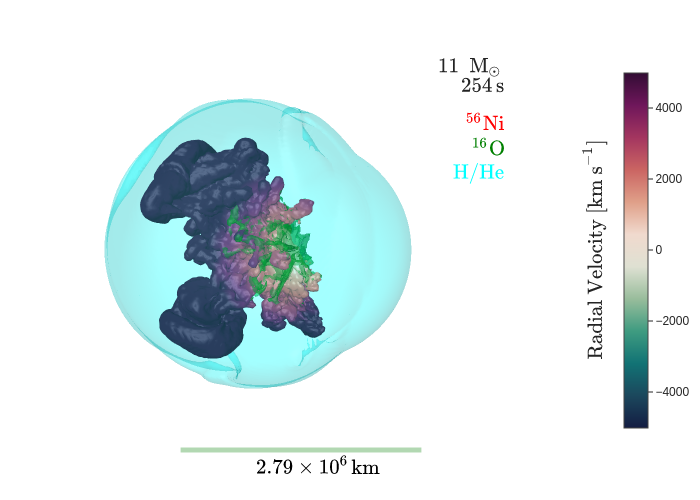}
    \includegraphics[width=0.49\textwidth]{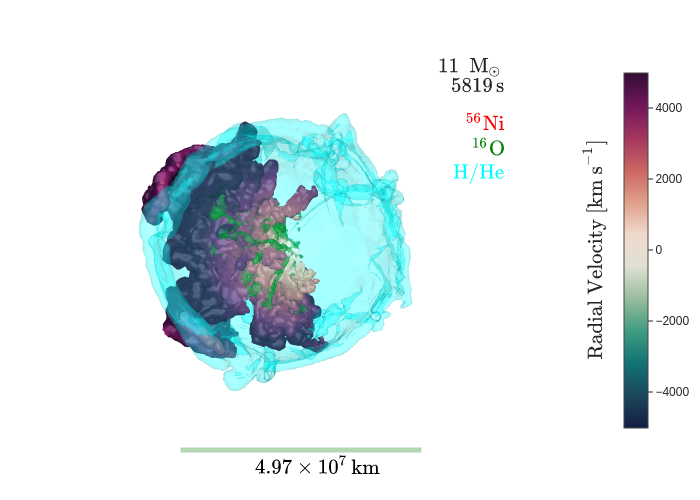}
    \includegraphics[width=0.49\textwidth]{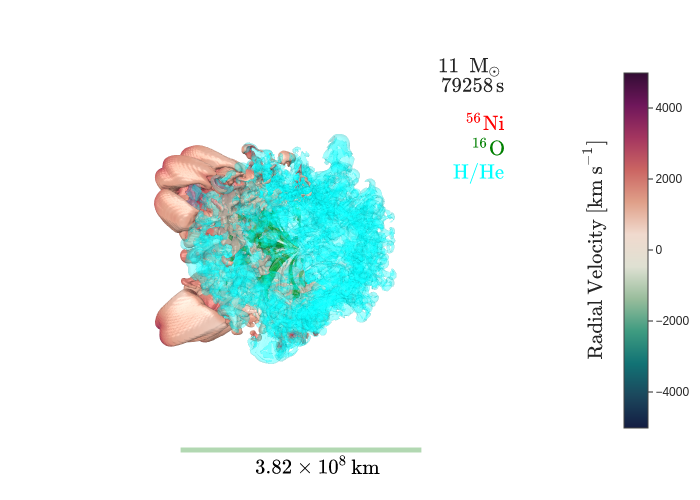}
 \includegraphics[width=0.49\textwidth]{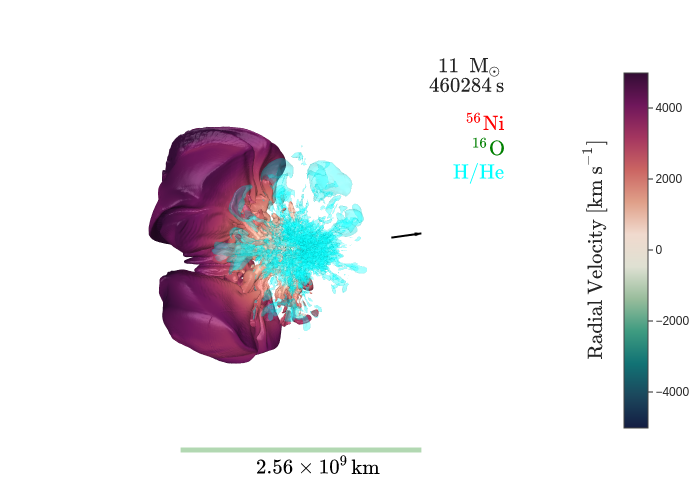}  
   \caption{Same as Fig.\,\ref{fig:9ni56}, but for the 11-M$_{\odot}$ progenitor.}
    \label{fig:11ni56}
\end{figure*}

\begin{figure*}
\centering
    \includegraphics[width=0.49\textwidth]{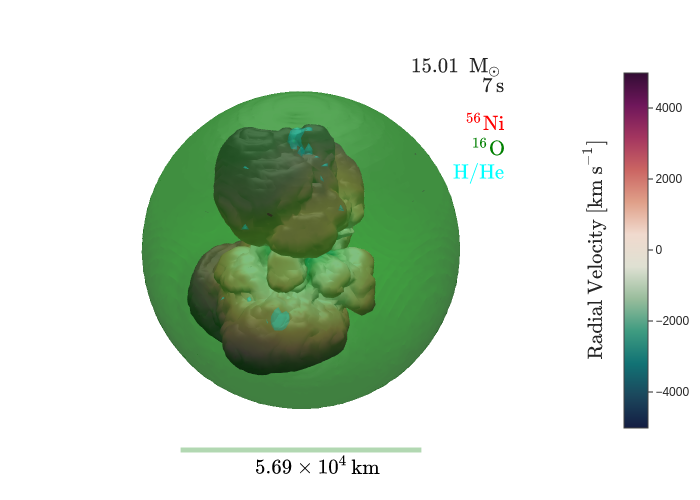}  
    \includegraphics[width=0.49\textwidth]{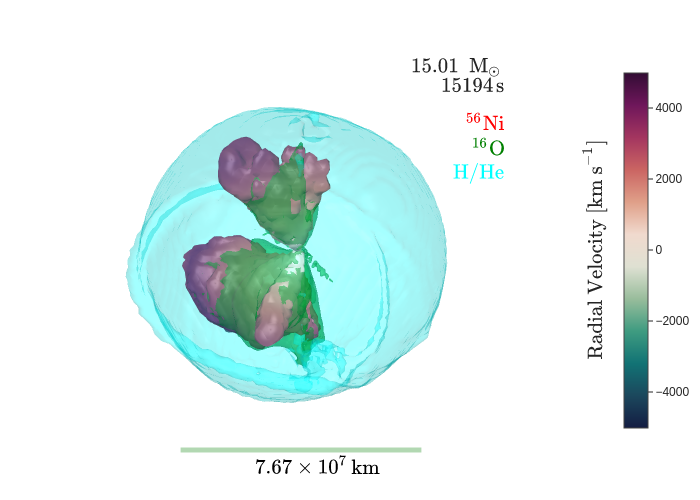}  
    \includegraphics[width=0.49\textwidth]{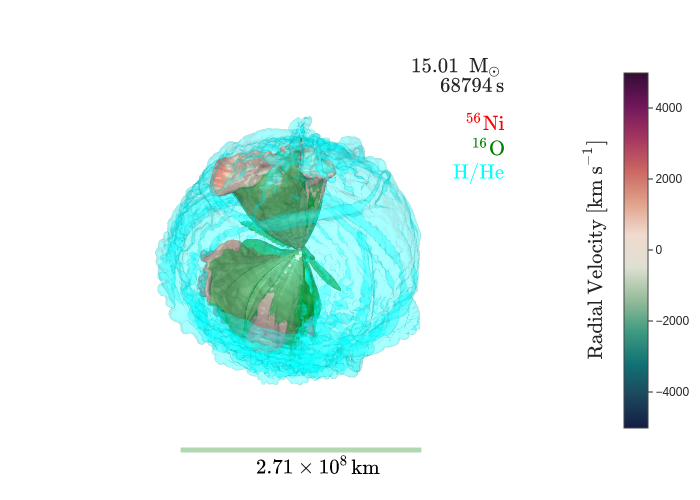}  
    \includegraphics[width=0.49\textwidth]{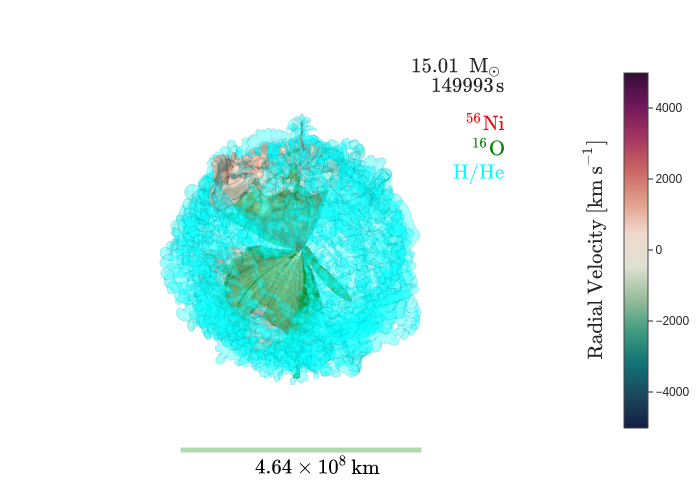}  
    \includegraphics[width=0.49\textwidth]{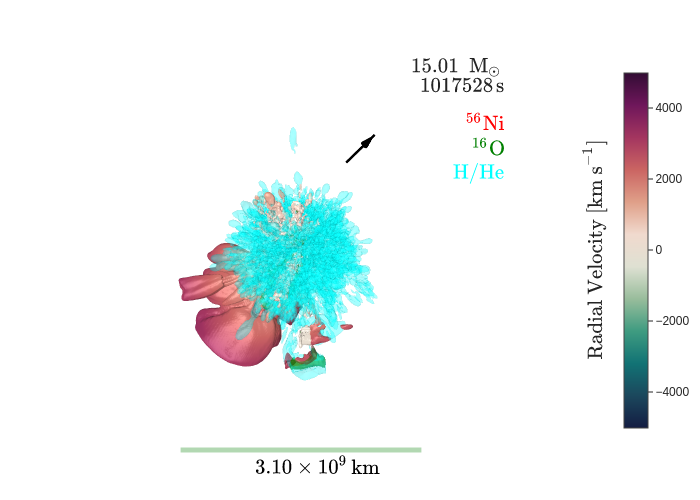}  
            
   \caption{Same as Fig.\,\ref{fig:9ni56}, but for the 15.01-M$_{\odot}$ progenitor.}
    \label{fig:15.01ni56}
\end{figure*}

\begin{figure*}
\centering
    \includegraphics[width=0.49\textwidth]{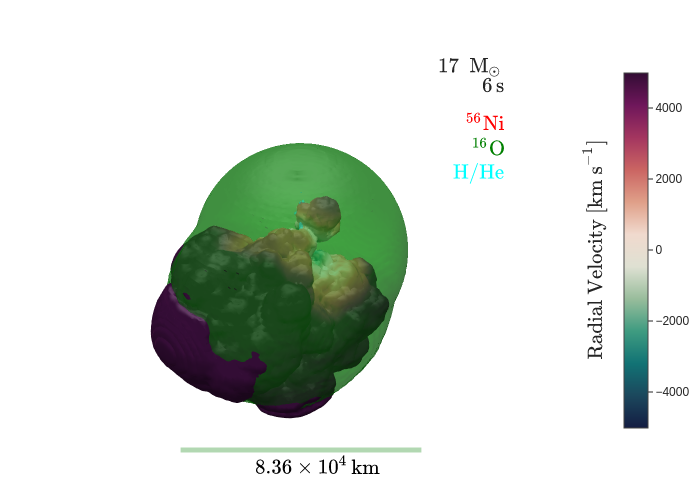}
    \includegraphics[width=0.49\textwidth]{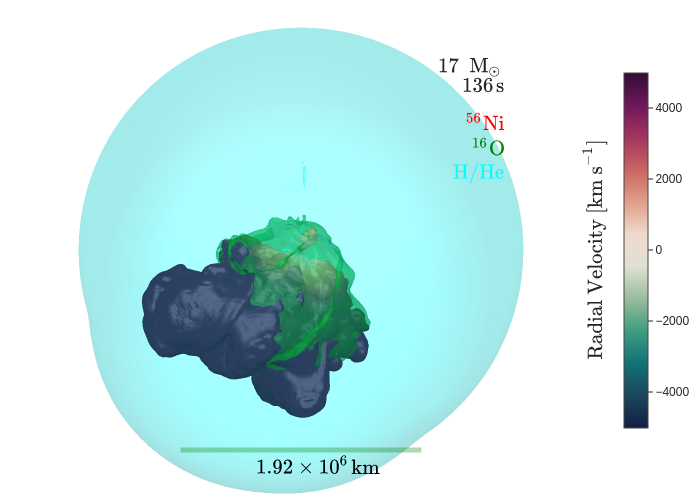}
     \includegraphics[width=0.49\textwidth]{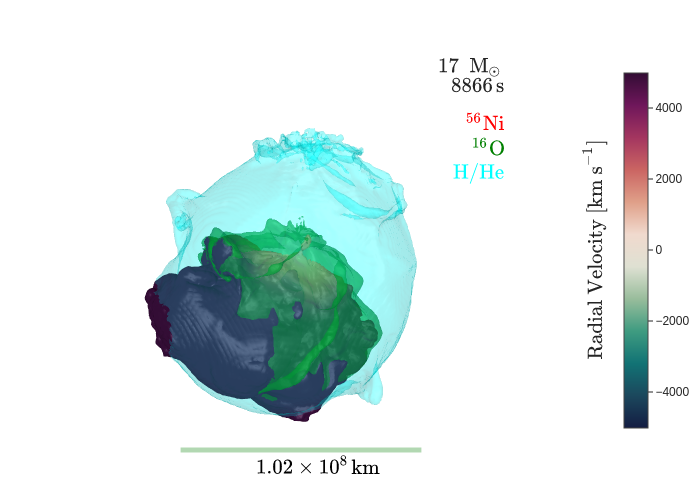}
    \includegraphics[width=0.49\textwidth]{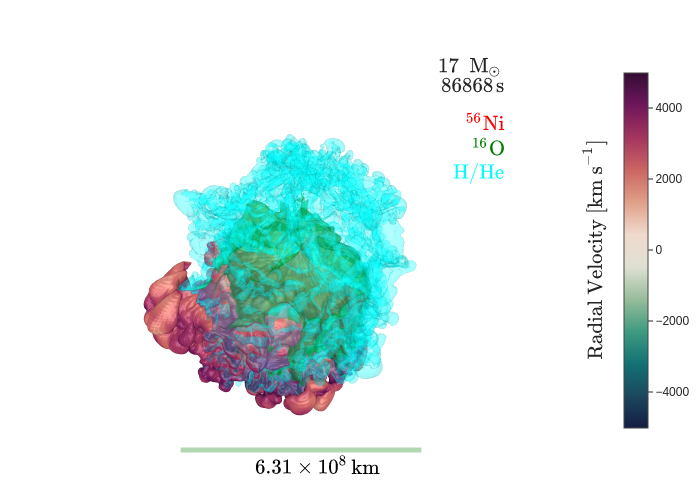}
    \includegraphics[width=0.49\textwidth]{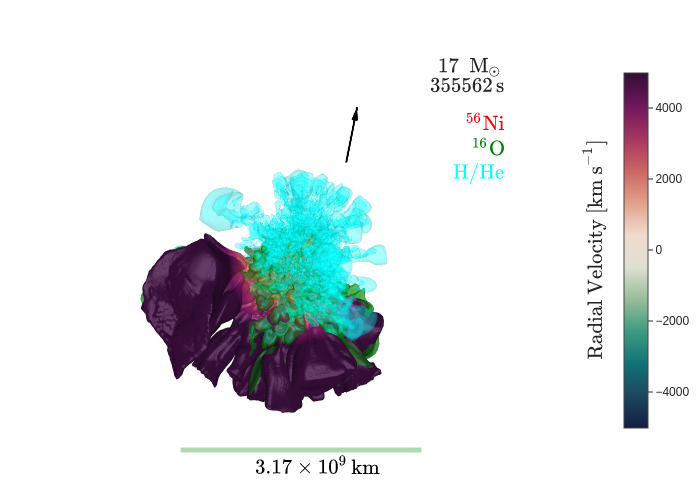}
   \caption{Same as Fig.\,\ref{fig:9ni56}, but for the 17-M$_{\odot}$ progenitor.}
    \label{fig:17ni56}
\end{figure*}

\begin{figure*}
\centering
    \includegraphics[width=0.49\textwidth]{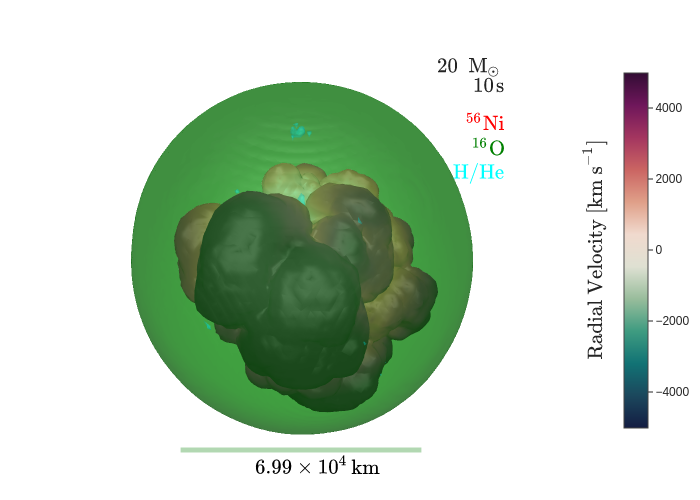}
    \includegraphics[width=0.49\textwidth]{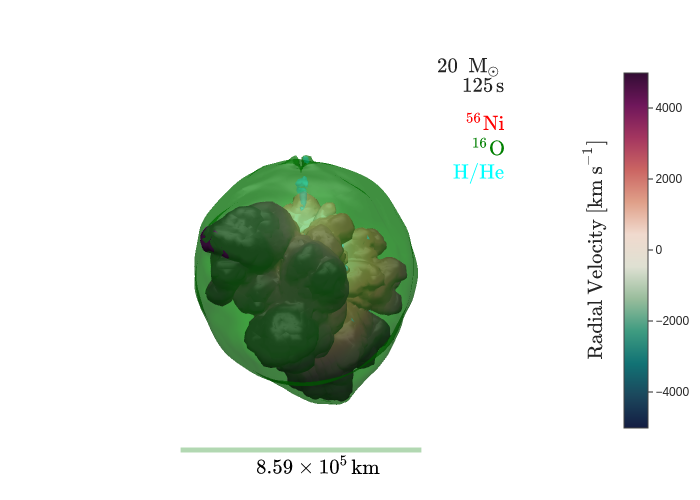}
    \includegraphics[width=0.49\textwidth]{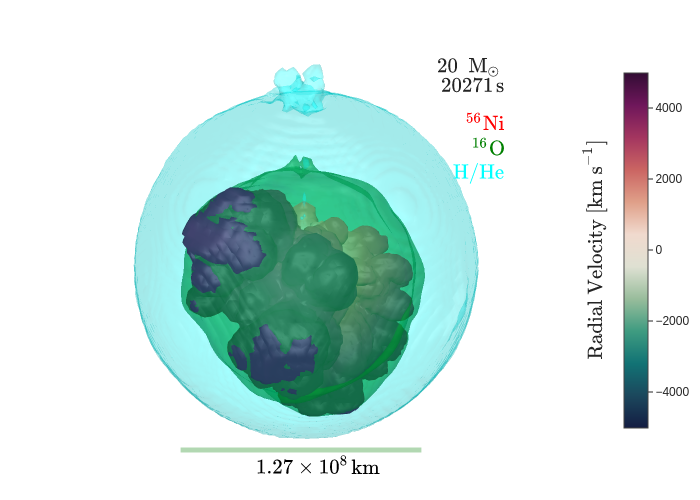}
    \includegraphics[width=0.49\textwidth]{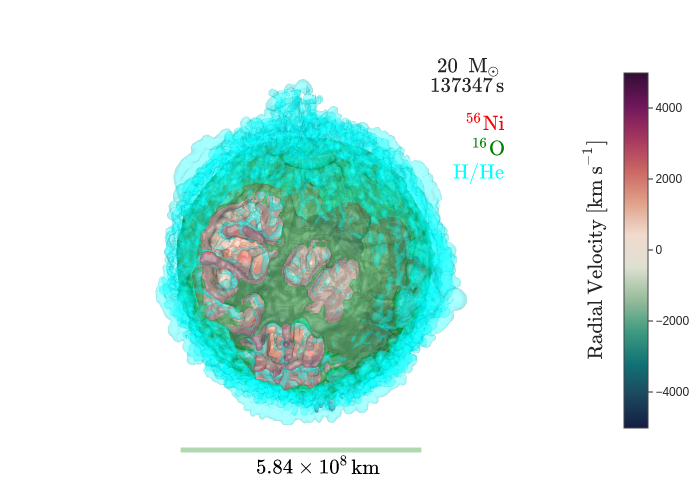}
    \includegraphics[width=0.49\textwidth]{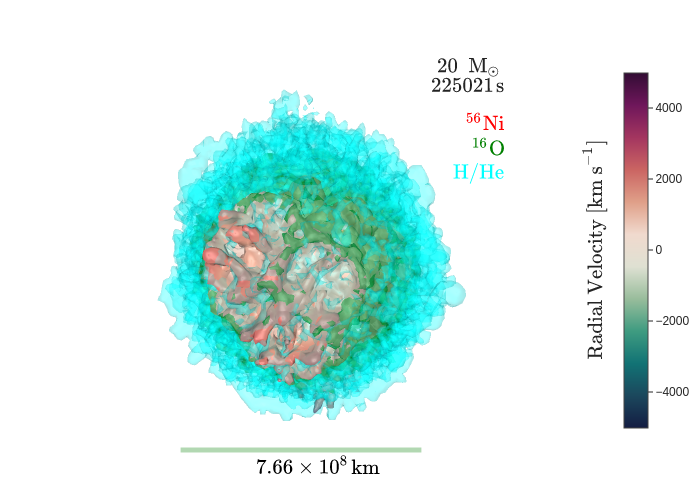}
    \includegraphics[width=0.49\textwidth]{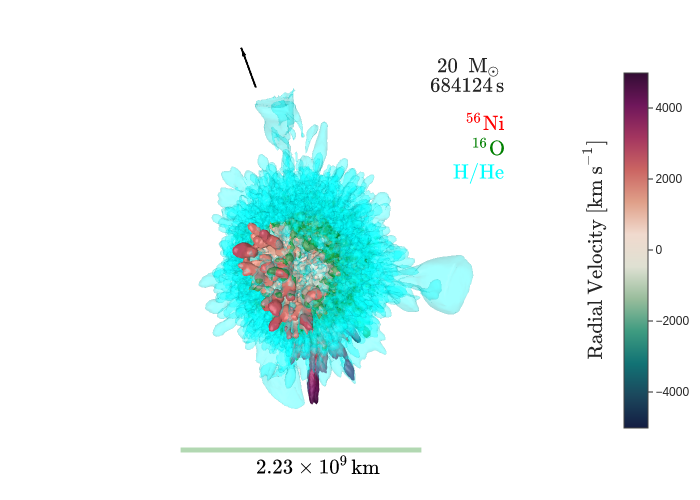}
   \caption{Same as Fig.\,\ref{fig:9ni56}, but for the 20-M$_{\odot}$ progenitor.}
    \label{fig:20ni56}
\end{figure*}

\begin{figure*}
\centering
    \includegraphics[width=0.49\textwidth]{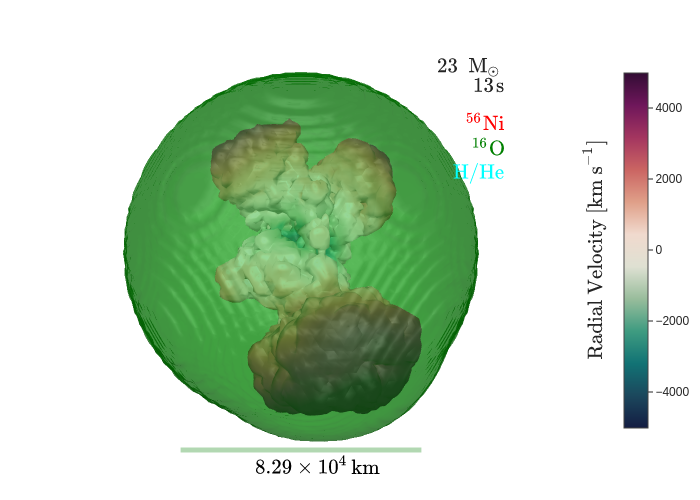}
        \includegraphics[width=0.49\textwidth]{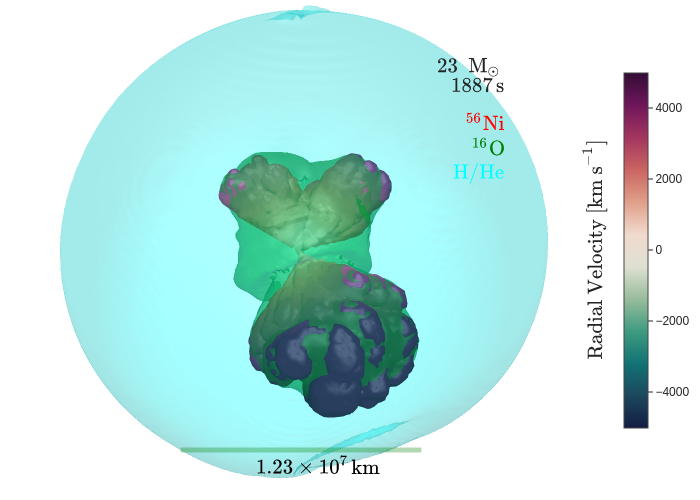}
    \includegraphics[width=0.49\textwidth]{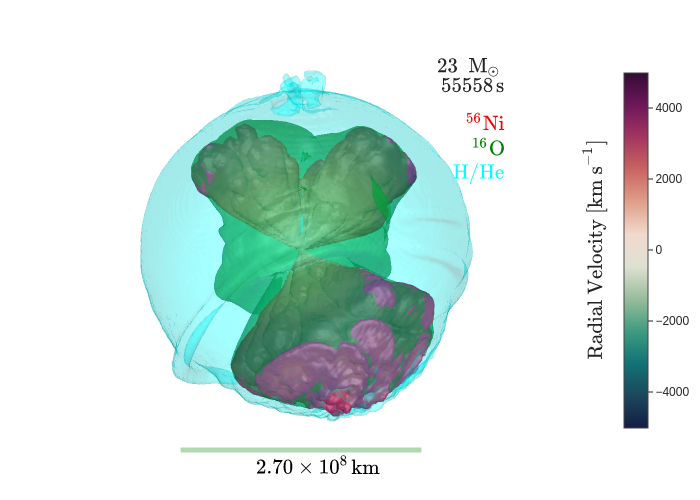}
     \includegraphics[width=0.49\textwidth]{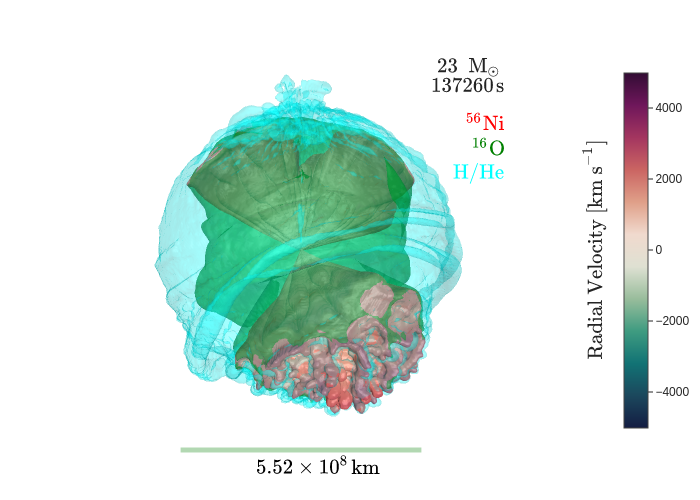}
\includegraphics[width=0.49\textwidth]{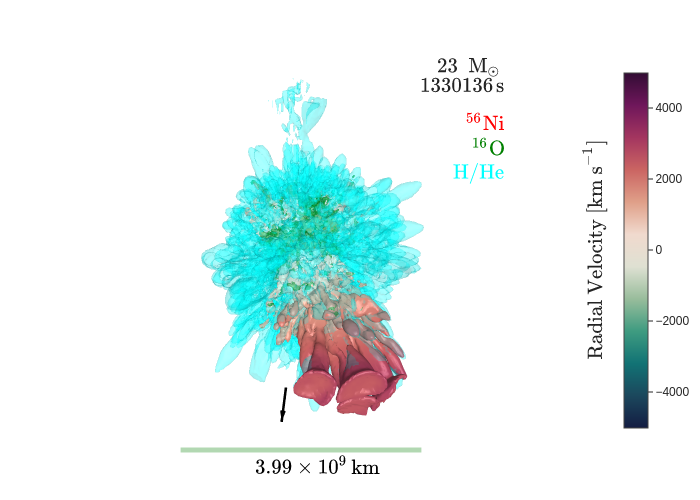}
   \caption{Same as Fig.\,\ref{fig:9ni56}, but for the 23-M$_{\odot}$ progenitor.}
    \label{fig:23ni56}
\end{figure*}

\begin{figure*}
\centering
    \includegraphics[width=0.49\textwidth]{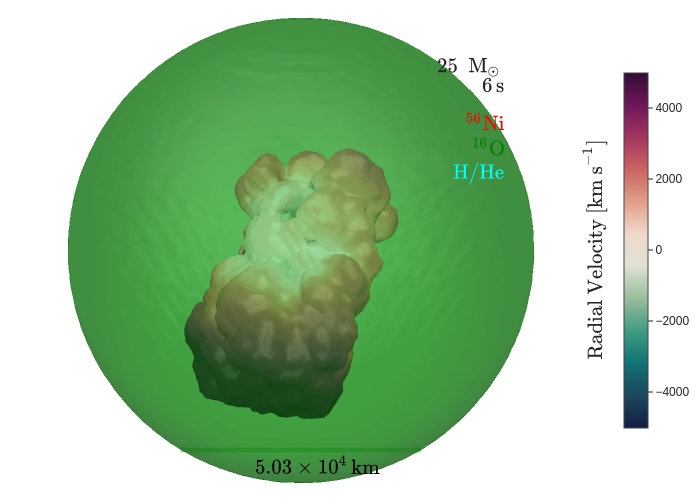}
    \includegraphics[width=0.49\textwidth]{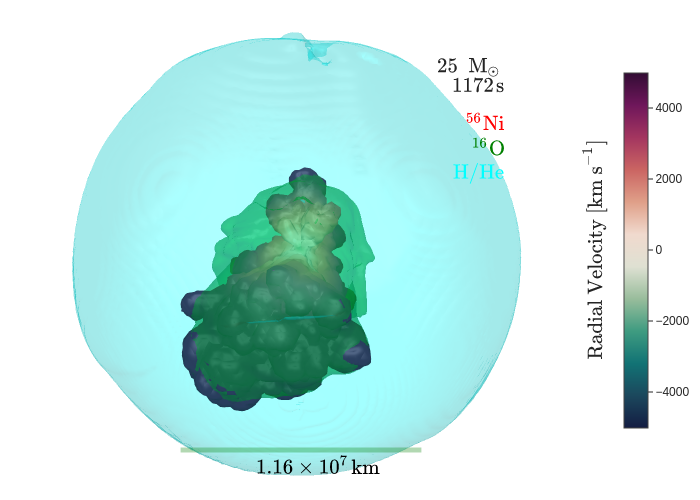}
    \includegraphics[width=0.49\textwidth]{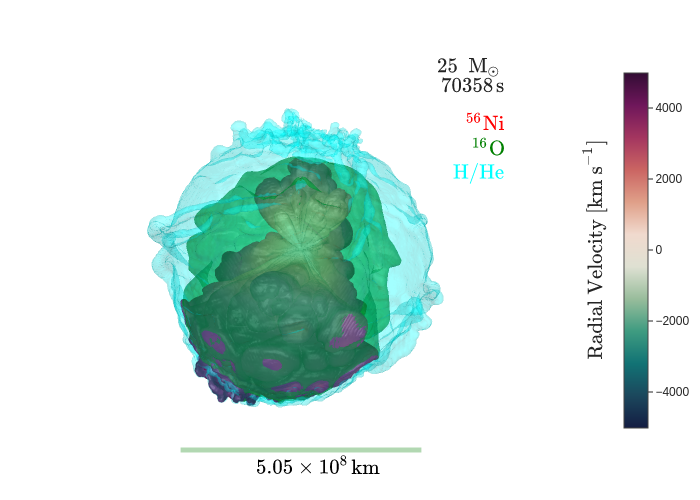}
    \includegraphics[width=0.49\textwidth]{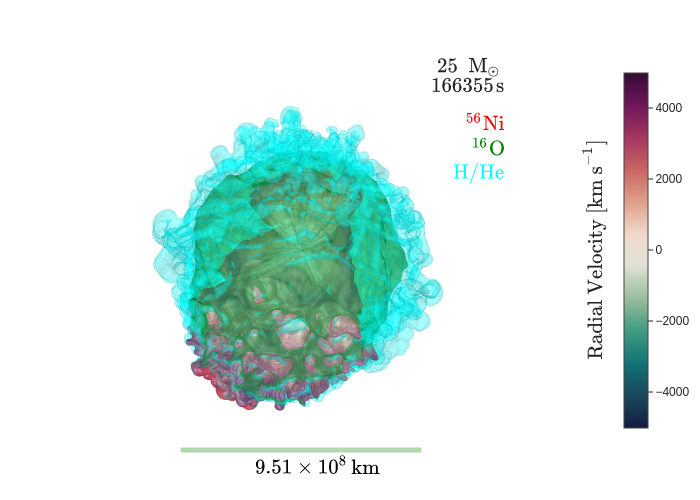}
    \includegraphics[width=0.49\textwidth]{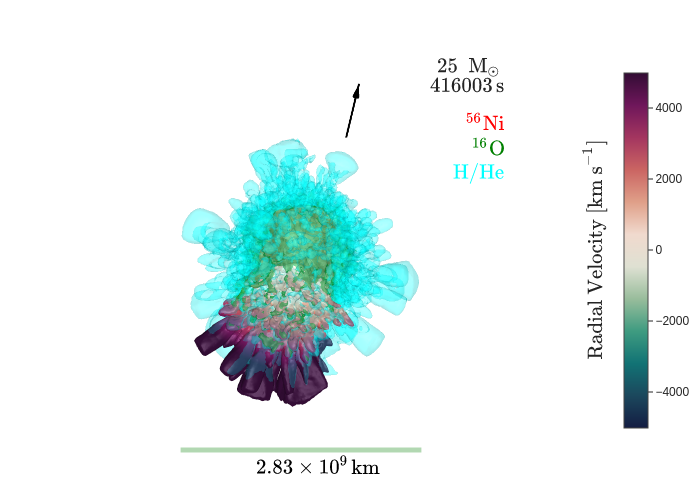}

   \caption{Same as Fig.\,\ref{fig:9ni56}, but for the 25-M$_{\odot}$ progenitor. Note again the emergence of $^{56}$Ni from the oxygen enveloping it in the first thousands of seconds. We see strong kick-ejecta anti-alignment.}
    \label{fig:25ni56}
\end{figure*}

We begin by describing the ejecta distribution from the early seconds of explosion through shock breakout from the RSG surface and continued until the ejecta expansion reaches homology. In Figs.\,\ref{fig:9ni56}$-$\ref{fig:25ni56}, we illustrate at various time snapshots the 1$\%$ isosurface of the $^{56}$Ni ejecta (red), the 30$\%$ isosurface of $^{16}$O (green), and the H/He interface (cyan) for the seven models we evolve here. Note that the latter two began as perfect spherical surfaces (beginning from a spherically-symmetric, 1D progenitor model). The first snapshots are near the end of the F{\sc{ornax}} simulations of neutrino-driven CCSNe out to several seconds; the final snapshots are following the emergence of shock breakout into its environment. We do not find a significant difference in our particular choice of the  $^{56}$Ni isosurface, which we vary from one to several percent. We choose 30$\%$ for the $^{16}$O isosurface to capture the C+O/He core and not an oxygen shell further out in the stellar envelope. Below, we discuss the structural evolution of the different nuclear compositions, emphasizing morphology and the preservation of asymmetries seeded in the first seconds by turbulent neutrino heating. 

\begin{itemize}
\item The 9-M$_{\odot}$ model $^{56}$Ni ejecta morphology (Fig.\,\ref{fig:9ni56}) has a moderately strong dipole moment, but with significant contributions from the quadrupolar and octupolar components. 
This is characteristic for lower mass progenitors, which generally produce less asymmetric explosions (lower low-$\ell$ angular power). At $\sim$2 s, the $^{56}$Ni isosurface has already mixed into the C/O interface. As the ejecta expand, asymmetries become more pronounced until the ejecta reach the H/He interface. 
We see the formation of mushrooming Rayleigh-Taylor instabilities (RTI) as the $^{56}$Ni interacts with the reverse shock.  The latter forms early and first propagates inward in mass (the shock front becomes wider in radial and mass coordinates) then in radius before converging on itself off-center (see also \citealt{sandoval}) and evolving into a forward-moving blast wave. 
After breakout, the $^{56}$Ni ejecta accelerate into a low density environment and result in distinct, elongated clumps of scales of $\sim$billions of km. We emphasize how aggressively outward the $^{56}$Ni is mixed, completely penetrating the oxygen ejecta.

\item The 11-M$_{\odot}$ model (Fig.\,\ref{fig:11ni56}) ejecta evince a strong dipolar geometry and much weaker higher-order modes. Again, as the $^{56}$Ni ejecta expand in the first hundreds to thousands of seconds, we see asymmetries become more pronounced. We see two lobes develop and, after interaction with the H/He interface, compress and decelerate. The advent of RTI at the interface is again marked by wrinkling of the $^{56}$Ni ejecta surface. These two lobes preserve their structure past shock breakout into a low-density CSM, accelerating into two large prominent features that again overwhelm the oxygen ejecta in outward mixing. The 11-M$_{\odot}$ model is exceptional in its excessive mixing. 

\item The 15.01-M$_{\odot}$ (Fig.\,\ref{fig:15.01ni56}) model ejecta are dominated by small-scale structure, with stronger quadrupolar and octupolar modes, and a weaker dipole mode comparable in strength to higher-order modes. However, within the first thousands of seconds, the ejecta dipolar and quadrupolar modes dominate. We see a later and weaker reverse shock with weaker $^{56}$Ni mixing. Little $^{56}$Ni ejecta is present exterior to the H/He interface after shock breakout.

\item The 17-M$_{\odot}$ model (Fig.\,\ref{fig:17ni56}) has a dominant dipole mode, similar to the 11-M$_{\odot}$, but with a weaker (still significant) quadrupole mode, and weaker contributions from higher $\ell$ modes. This high-energy (1 Bethe (B) [$\equiv$10$^{51}$ erg] explosion energy) model is also significant in its strong $^{56}$Ni mixing outward, and its strongly dipolar mode is preserved beyond shock breakout.

\item Model 20-M$_{\odot}$ (Fig.\,\ref{fig:20ni56}) has a strong dipole mode in the ejecta but maintains strong smaller-scales modes traversing to the higher $l$ cascade. At breakout, this explosion has one of the most spherical geometries and, incidentally, less $^{56}$Ni mixing. 

\item The 23-M$_{\odot}$ (Fig.\,\ref{fig:23ni56}) ejecta morphology is marked by strong $l=2-4$ contributions dominating over the dipole mode. This model is a moderate/weak energy explosion and yields the least mixed $^{56}$Ni of all our models. We see the formation of three large-scale plumes preserved through shock-breakout, with a dominant plume in the southern hemisphere. 

\item Lastly, the 25-M$_{\odot}$ (Fig.\,\ref{fig:25ni56}) model $^{56}$Ni ejecta have a strong quadrupolar mode dominating the dipole mode, and additional appreciable contributions from higher-order modes. The dipole mode begins to dominate the $^{56}$Ni ejecta in the first 100s of seconds until the ejecta encounters the reverse shock at the H/He interface, at which point the quadrupolar mode overtakes the dipole, with multiple large lobes of $^{56}$Ni ejecta emerging. Despite having a similar explosion energy to the 17-M$_{\odot}$ progenitor, this model, like the 20-M$_{\odot}$ model, has much less $^{56}$Ni mixed outward. We discuss the reasons in \S\,\ref{subsec:mix}. 

\end{itemize}

As seen on the last still in Figs.\,\ref{fig:9ni56}$-$\ref{fig:25ni56}, the remnant kick direction is anti-aligned with the $^{56}$Ni distribution for all our models, except the 9- and 23-M$_{\odot}$ progenitors. We expect anti-alignment from momentum conservation $-$ the neutron star (NS) is kicked in the opposite direction of the explosion ejecta (e.g. \citealt{kick}, but also     \citealt{2024Ap&SS.369...80J}). The former, as a lower mass progenitor, explodes more symmetrically (and with a weaker energy); the latter has strong multipolar modes in its explosion morphology from the first seconds, including two plumes expanding along the northern hemisphere. The 23-M$_{\odot}$ model is also an outlier as a low-energy, weakly-asymmetric explosion resulting from a more massive progenitor.\footnote{The 23-M$_{\odot}$ progenitor also forms a $\sim$5\,M$_{\odot}$ black hole after several hours \citep{2025ApJ...987..164B}. The core can continue to accrete mass and fatten up while the explosion continues to eject material including $^{56}$Ni, and this amount is reflected in Fig.\,\ref{fig:23ni56}.} Comparing to Fig.\,\ref{fig:mix}, we see indeed both these models have a weak ejecta dipole. This is also illustrated in Fig.\,17 in \cite{burrows2023} $-$ both these models have weak kicks, smaller ejecta and energy dipoles, and smaller explosion energies.\footnote{As is generically the case in CCSNe simulations, the core is held fixed on the simulation grid and the kick is calculated as integral over the neutron star surface.}

Furthermore, in both models the neutrino contribution to the total kick, which evinces a weaker correlation with the ejecta morphology than the matter contribution to the kick \citep{kick}, is significant and reflects weaker large-scale explosion asymmetries in the first seconds which would drive a larger matter kick. Thus, measurements of NS kicks anti-aligned with imaged $^{56}$Ni distributions would be an indirect confirmation of neutrino-heating as the mechanism for CCSNe, for which natal kick alignments persist through fallback accretion and shock breakout (\citealt{2022ApJ...926....9J}, but see also \citealt{2023MNRAS.526.2880M,chan}). 

\subsection{Structural Correlation with Time}

Having discussed the differences between the ejecta properties of the models studied here, we can summarize a general picture of hydrodynamic evolution following the blast propagation through shock breakout. We see to varying degrees the penetration of $^{56}$Ni beyond the overlaying oxygen surface via multi-scale ejecta structures for all the models.
Interaction with the H/He interface results in the small-scale wrinkling of the $^{56}$Ni isosurface through RTI.  

We first explore for red supergiant explosions how much of the seed structure of the ejecta imprinted in the first seconds of the neutrino-driven explosion is preserved over time. We investigate the evolution of the dipole direction of the $^{56}$Ni isosurface from the first seconds of explosion to shock breakout. The $^{56}$Ni structure in models 15.01- and 23-M$_{\odot}$ (and to a smaller degree model 25-M$_{\odot}$) is dominated by higher-order moments, and, hence, these models exhibit slight drift in the dipole direction. Regardless, the initial pointing of the $^{56}$Ni surface is remarkably well-preserved in time from seconds to days for all models studied here.

To capture the degree to which global measures of asymmetry are preserved, we explore the self-correlation (see \S\ref{self}) of the $^{56}$Ni ejecta structure at any given time with its structure at the time of mapping into FLASH. 
Initially, the self-correlation of the ejecta morphology decreases with expansion in the first 1000s of seconds. This is characterized by an acceleration of the shock and the ejecta, with small-scale seed instabilities magnified as the blast progresses\footnote{Generally, fallback might also affect the early morphology of the ejecta on 1000s of seconds.}. The shock encounters the H/He interface first (after a model-dependent $\sim$100s of seconds) 
and decelerates.  We see the subsequent development of a reverse shock that first propagates inward in mass, then in radius. The $^{56}$Ni ejecta, trailing the shock (the extent is model-dependent), will encounter the H/He interface later ($\sim$1000s of seconds) and decelerate in turn. Interaction of the $^{56}$Ni ejecta with the reverse shock forming at the H/He interface for RSGs will seed the necessary conditions for RTI, which both mixes the $^{56}$Ni outwards and imprints small-scale structure onto its surface. The self-correlation of the ejecta increases as the ejecta `flattens' into the H/He interface. 

We now briefly discuss the evolution of the {\it normalized} $^{56}$Ni surface area as associated with the formation of small-scale clumps (see \S\ref{sec:clumping}). Upon the ejecta reaching the H/He interface, this normalized area decreases in models with a strong dipole moment characterizing the ejecta morphology, but increases in models with a strong multipolar ejecta morphology. A dipolar morphology for the $^{56}$Ni surface indicates that much of the $^{56}$Ni is concentrated along a particular direction, which decelerates at the H/He interface. However, multiple smaller modes at different radial scales in models with significant multipolar ejecta do not experience the same effect $-$ when one clump decelerates into the interface, another clump deeper in may still be accelerating and not yet at the interface. The H/He interface sets the stage for RTI formation, and the self-correlation of the ejecta again decreases as the $^{56}$Ni ejecta surface fragments. The 9-M$_{\odot}$ model $^{56}$Ni clumping ratio reaches a factor of roughly five. The 20 M$_{\odot}$ clumping ratio also becomes large (a factor of several) because it maintains a more spherical ejecta geometry with a surface mottled by hydrodynamic instabilities. For this model, the $^{56}$Ni ejecta have a large volume filling fraction, but a small mixing fraction.  On the other hand, the clumping fraction of the 17-M$_{\odot}$ model dips below one just before RTI formation because of this model's strong dipolar structure. 


RTI result in small-scale wrinkling of the $^{56}$Ni isosurface into clumps, increasing the area of the $^{56}$Ni isosurface relative to the angle-averaged $^{56}$Ni isosurface area. A larger dipole fraction leads to more efficient mixing of the ejecta through the base of the hydrogen envelope (see also Fig.\,\ref{fig:mix}). Development of RTI requires pressure and density gradients anti-aligned developed by subsequent acceleration and deceleration phases \citep{1989ApJ...341L..63A}. For our series of RSG models, the shock accelerates down the density gradient (a negative gradient in $\rho$r$^{3}$) in the outer helium envelope, where the helium mass fraction dips and the hydrogen mass fraction rises just interior to the original progenitor H/He interface, where the radial gradient of $\rho$r$^{3}$ increases. Thus, there is a  limited radial interval for acceleration then deceleration of the shock front. 
At the same time, the inertial $^{56}$Ni ejecta coasts at roughly a constant velocity with time.
The 9-M$_{\odot}$ experiences almost continuous growth of small-scale structure, despite lacking a coasting period, because of its low envelope mass. In contrast, the 17-M$_{\odot}$ progenitor features an early and extended coasting duration for the $^{56}$Ni ejecta until they encounter the H/He interface. The $^{56}$Ni clumping ratio mimics this `coasting' behavior seen in the $^{56}$Ni ejecta velocity, with both remaining nearly uniform until interaction with the H/He interface (see. e.g., left panel in Figure 5 in \citealt{vartanyan2025}).

For more massive models, the helium envelope lies further from the core and is smaller in radial width. 
Correspondingly, the region where the $^{56}$Ni can coast is smaller and further out, leading to delayed, weaker acceleration, RTI formation, and, hence, reduced $^{56}$Ni clumping and mixing.
We speculate that the extent of the envelope where the ejecta can accelerate, corresponding to the region where the envelope transitions from helium to hydrogen (interior to the H/He interface), may be crucial in establishing the extent of mixing.

The development of neutrino-driven turbulence at small scales and early times, has a certain stochastic character.  These small-scale structures may later coalesce into dominant dipolar distributions \citep{vartanyan2018b}, or bipolar and higher-order distributions \citep{burrows2024}.  Adding inner perturbations, either artificially to the 1D spherically-symmetric progenitors \citep{muller_janka_pert,burrows2018} or through the use of 3D initial conditions \citep{vartanyan2022} will also slightly alter the initial formation of structures; the direction of explosion may correlate with the structure of pre-collapse perturbations. With this caveat, however, we still see a robust correlation between the morphology of the ejecta and the explosion energy; the explosion energy also correlates with the progenitor properties, particularly with the core compactness \citep{burrows2024}. Hence, the explosion energy (for a fixed implementation of neutrino physics) is a reflection of the progenitor structure. Thus, while the details of the early-time turbulent structure is a sensitively stochastic process, we still expect, for the same progenitor, that the explosion energy will remain relatively consistent and the explosion morphology will as well, and so we expect the morphologies here to be physically meaningful. 

Our conclusion is \textit{not} that a particular RSG progenitor will leave behind a particular ejecta morphology (CCNSe progenitor structures show significant differences between different codes, even for the same mass, and identifying the final structure of a particular stellar profile is an unsolved problem). Rather, we find correlations between progenitor profiles (with whatever mass they correspond to relegated in some degree to a dummy index) and the ejecta morphology, while recognizing that capturing the range of explosion outcome distributions for a given CCSN progenitor with mild variations in the simulation setup is crucial.

\subsection{$^{56}$Ni Mixing and Ejecta Velocities}\label{subsec:mix}
\begin{figure*}
\centering
    \includegraphics[width=0.58\textwidth]{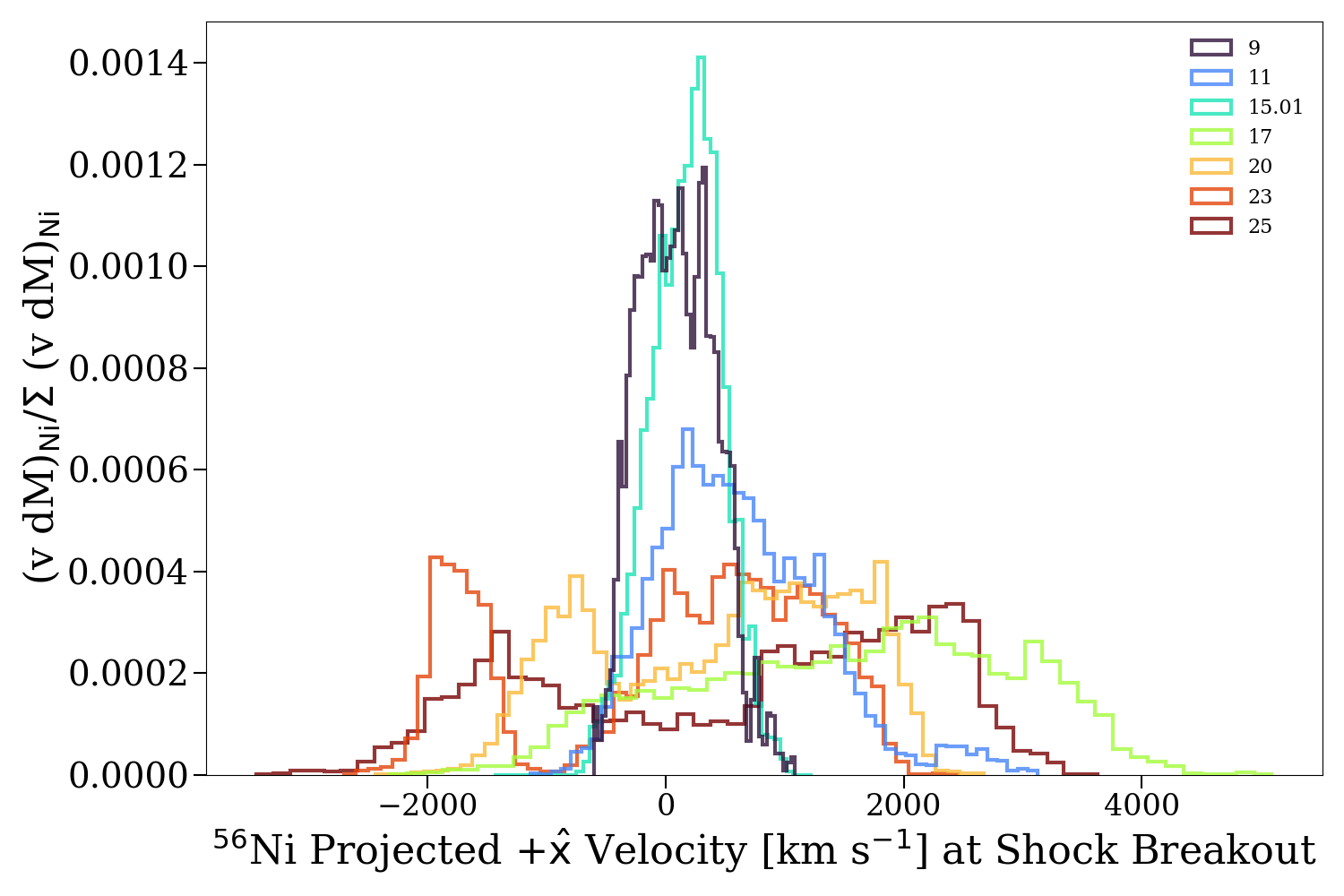}
    \includegraphics[width=0.58\textwidth]{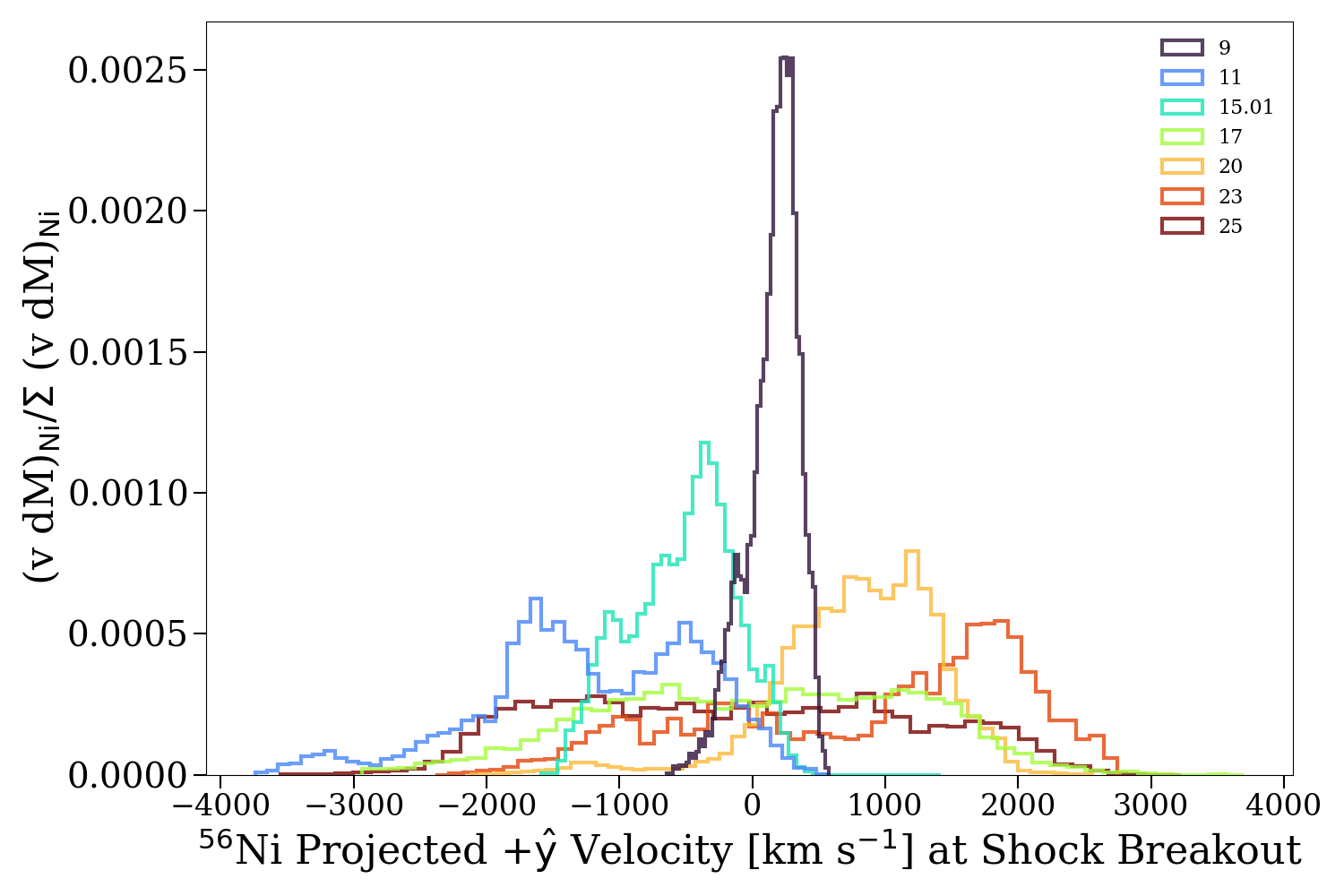}
    \includegraphics[width=0.58\textwidth]{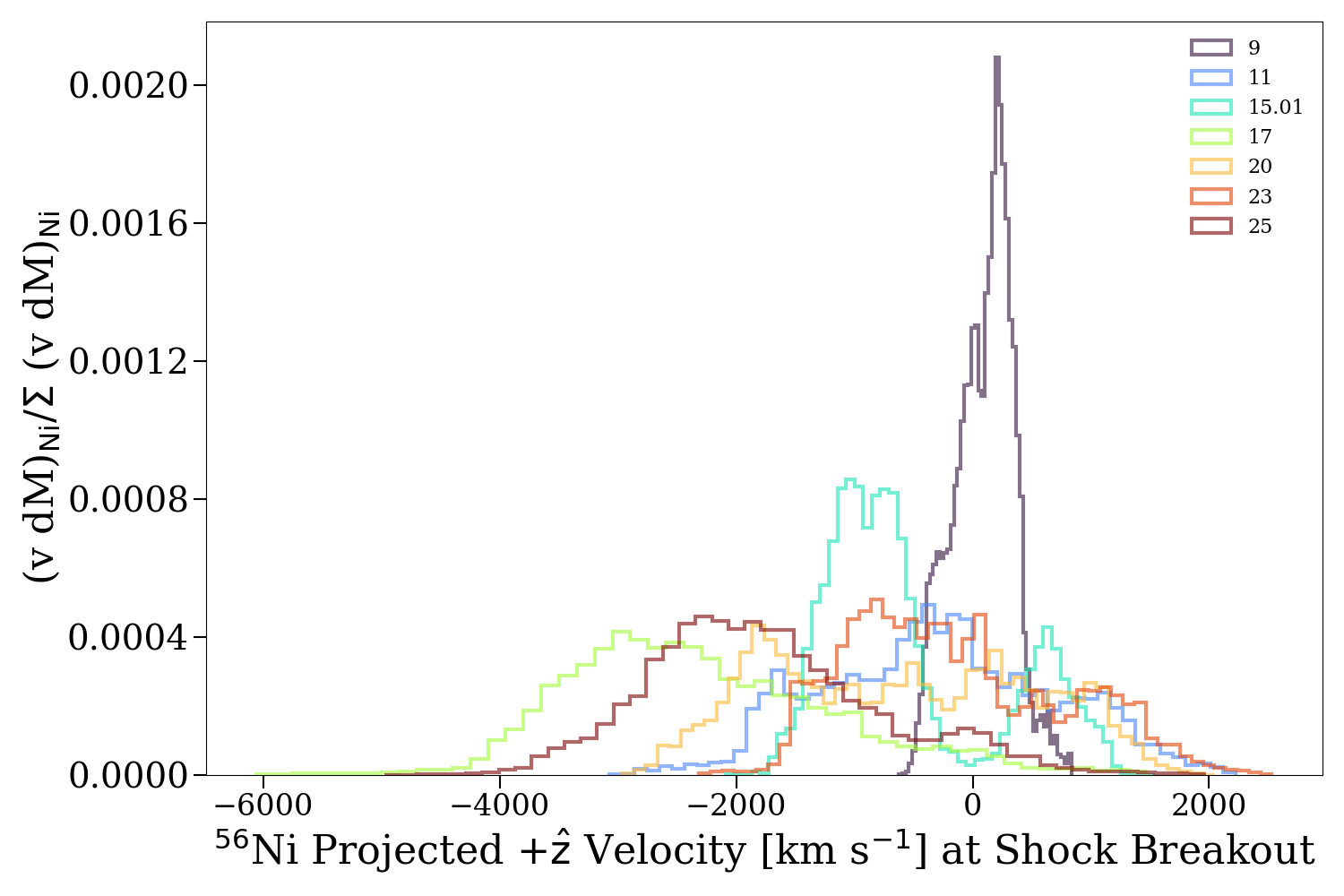}
   \caption{Mass-weighted $^{56}$Ni velocity at shock breakout, projected for viewing along the positive x,y, and z-directions, for our models. The sum of the area under the histogram is normalized to one. Note that the 17-M$_{\odot}$ has the highest $^{56}$Ni breakout velocity, and the 11-M$_{\odot}$ one of the highest maximum velocities in its $^{56}$Ni tail, despite not having a particularly high explosion energy. The ejecta velocity distribution reveals the clumped ejecta morphology, and we do not see shellular isotropic structure.}
    \label{fig:velx}
\end{figure*}

\begin{figure*}
\centering
    \includegraphics[width=0.45\textwidth]{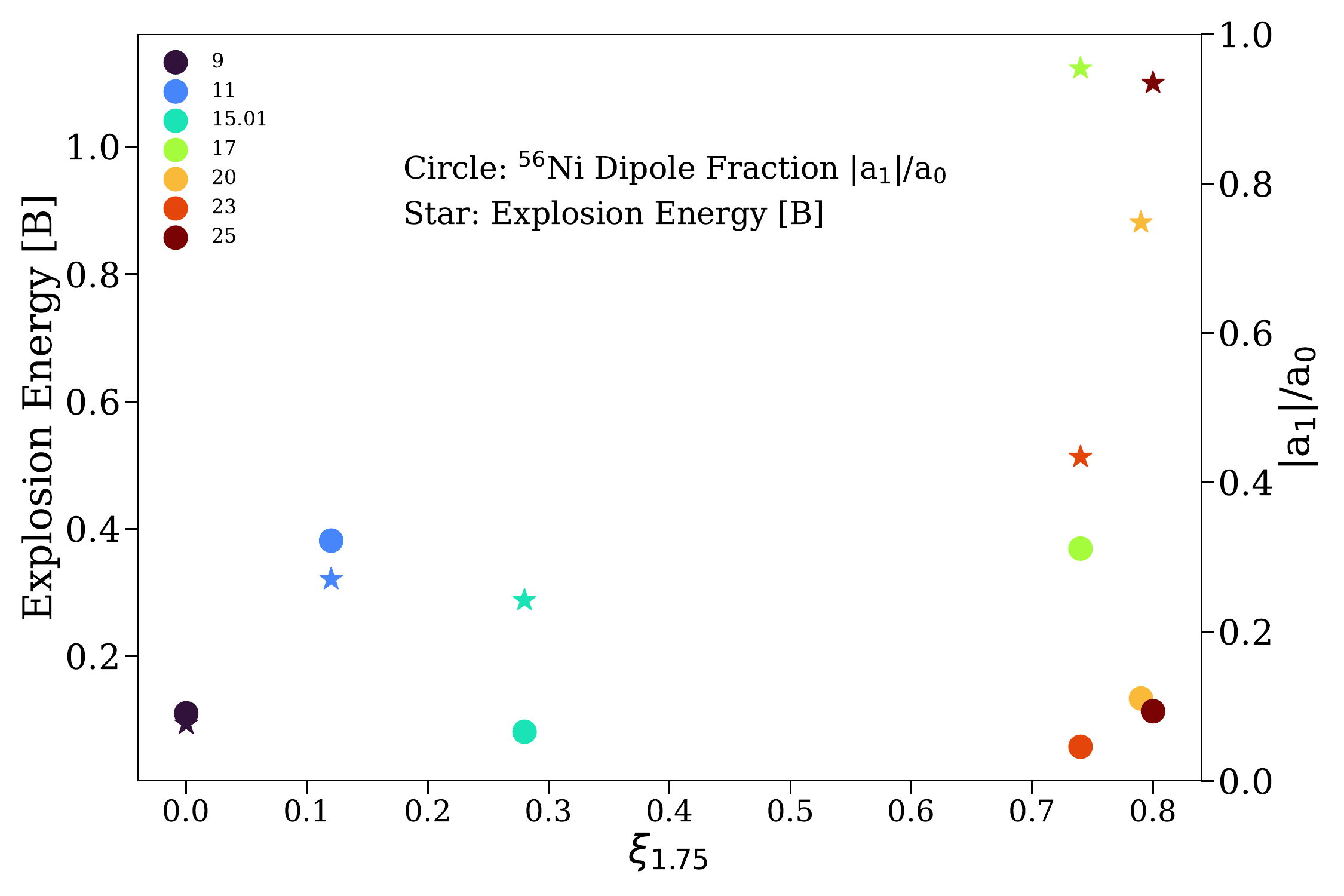}    
    \includegraphics[width=0.45\textwidth]{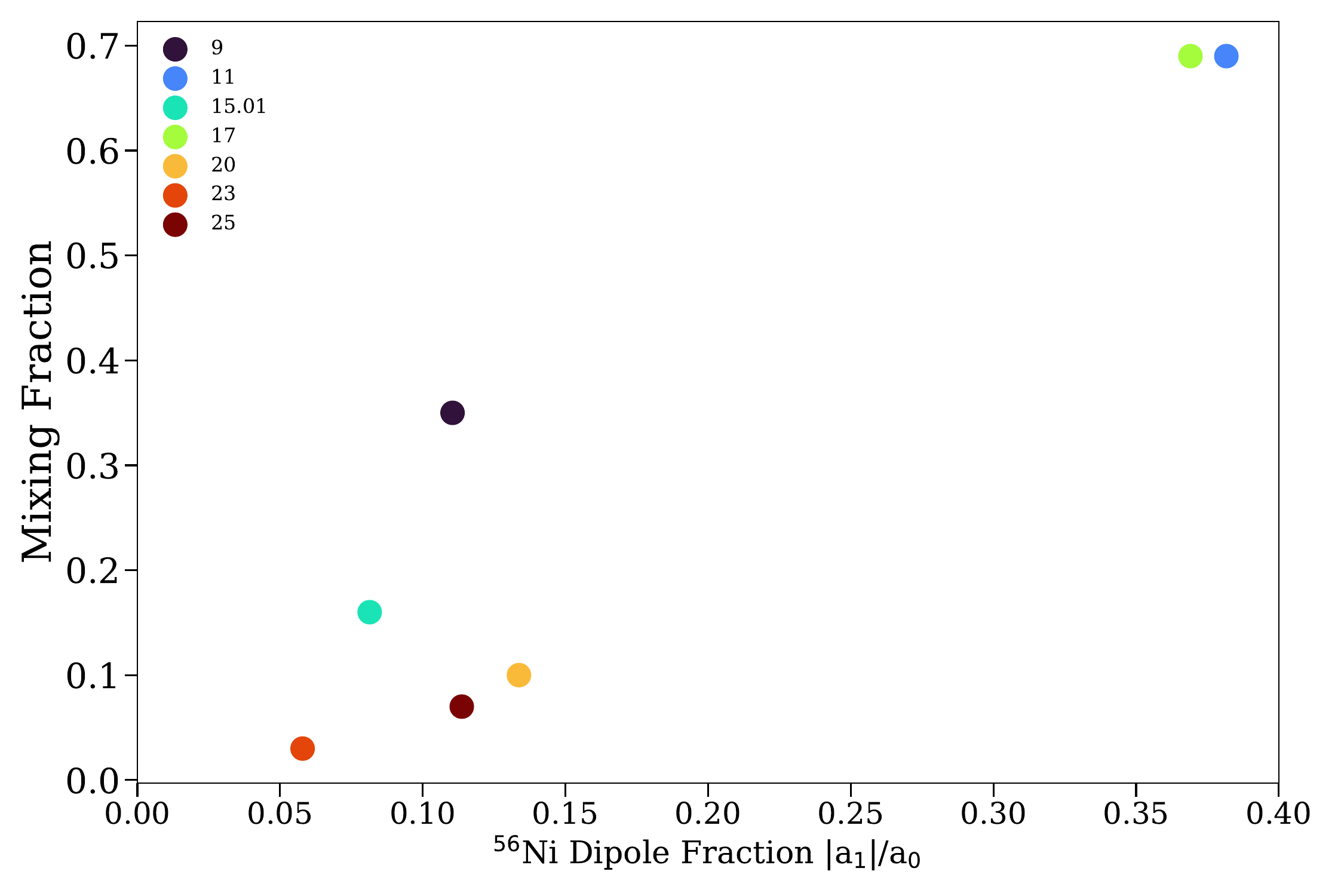}
      \includegraphics[width=0.45\textwidth]{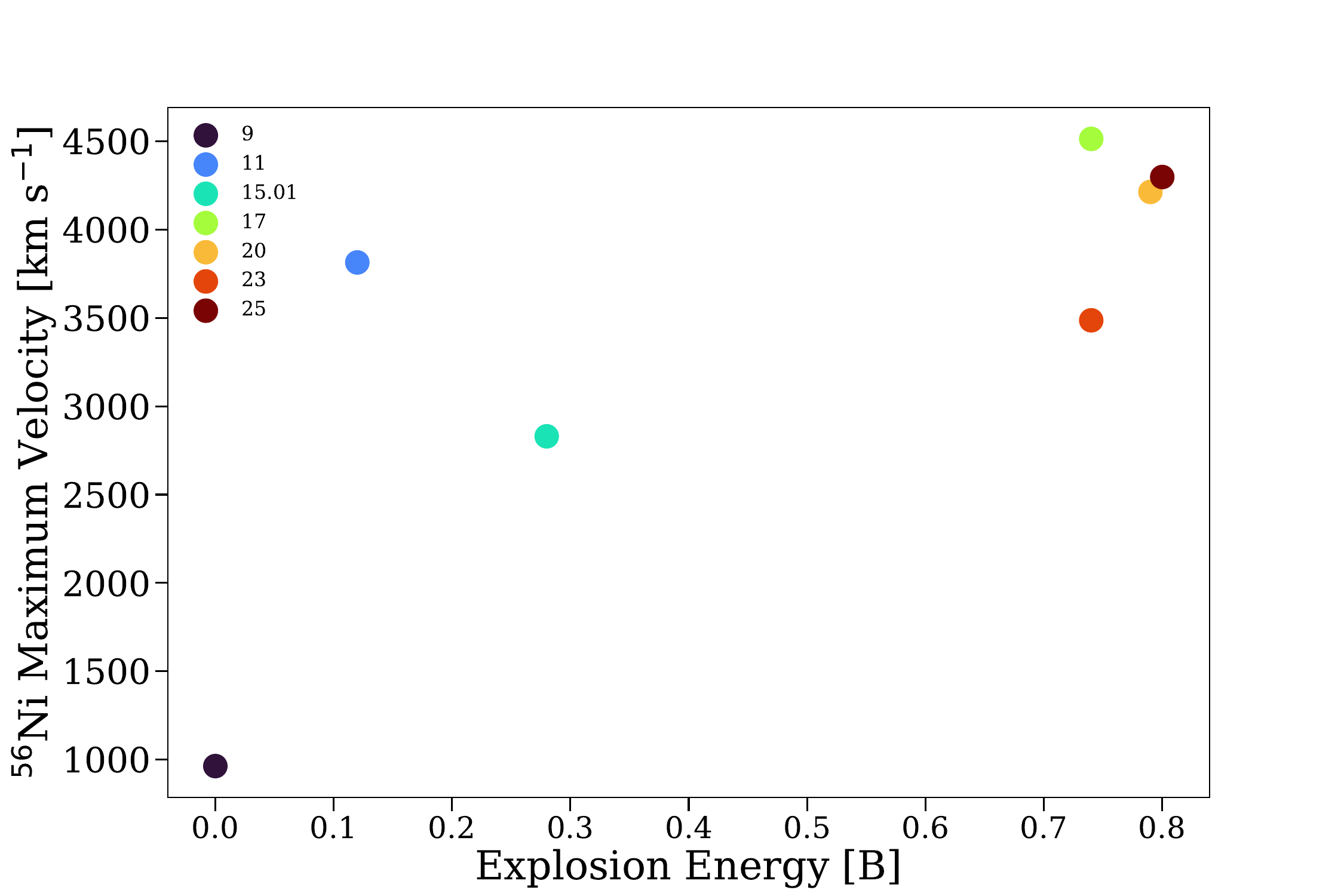}
    \includegraphics[width=0.45\textwidth]{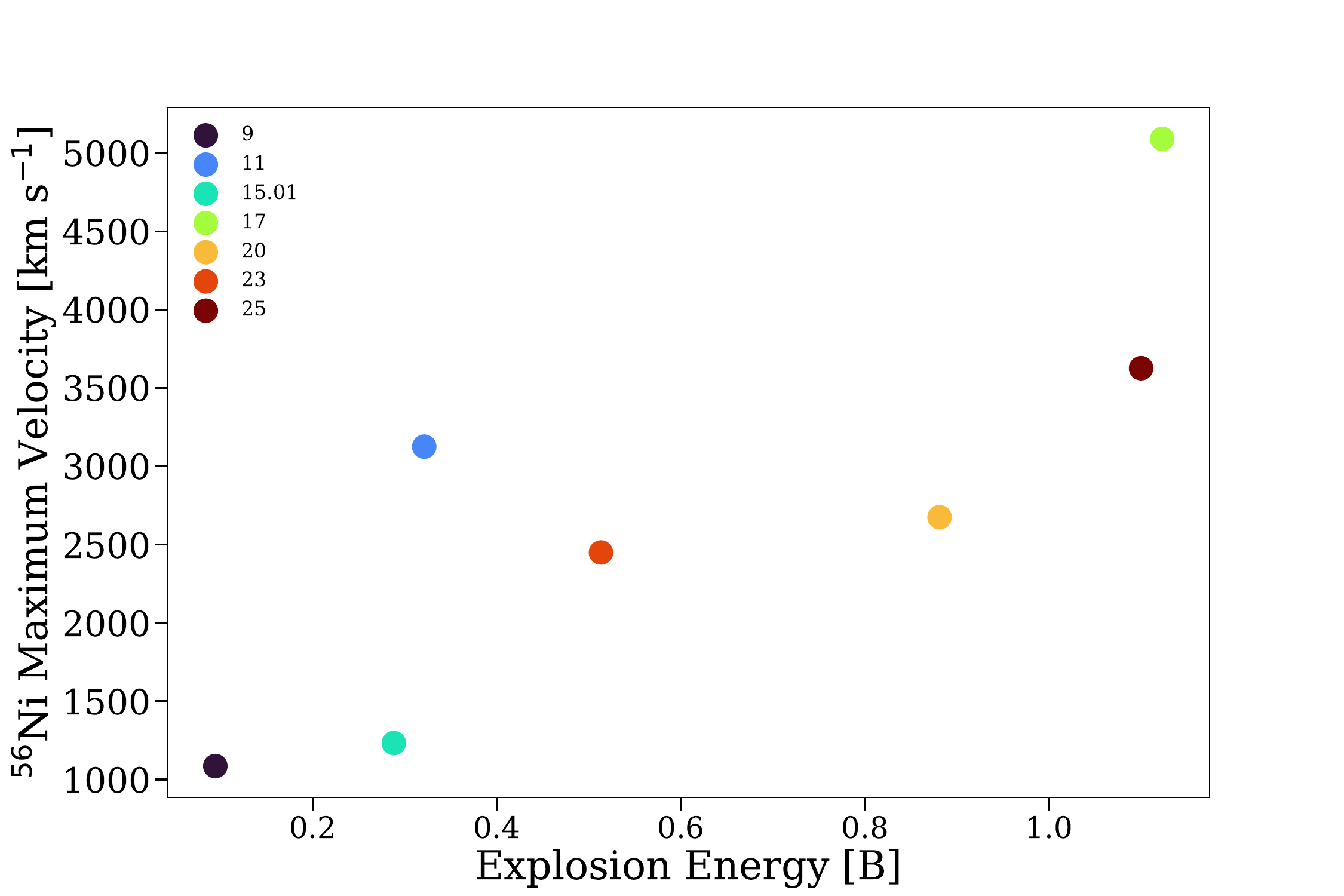}
   \caption{\textbf{Top left}: Explosion energy (star, in Bethe) and dipole fraction (circle) of the 3$\%$ $^{56}$Ni isosurface at the time of mapping to FLASH plotted against progenitor core compactness, 
   $\xi_{1.75}$ at 1.75 M$_{\odot}$. Compare to Fig.\,9 in \protect{\citealt{burrows2023}}, which shows the ejecta mass dipole. Note that the 11-M$_{\odot}$ and 17-M$_{\odot}$ represent local maxima in both explosion energy and dipole strength within their compactness neighborhoods. \textbf{Top right}:  We plot the fraction of $^{56}$Ni mixed into the hydrogen envelope against the  dipole fraction of the 3\% $^{56}$Ni isosurface at the end of the F{\sc{ornax}} simulation, at mapping into FLASH. $^{56}$Ni that is more asymmetrically produced by explosive nucleosynthesis through the end of the neutrino-driven explosion is more heavily mixed outwards. \textbf{Bottom:} We see a strong correlation of both the maximum $^{56}$Ni velocity and the spread in velocity (defined as the width at 10$\%$ of maximum) with core compactness and explosion energy.  See text for discussion.}
    \label{fig:mix}
\end{figure*}

We continue our discussion by exploring the progenitor dependence of the $^{56}$Ni mixed fraction and ejecta velocity, with some of our results summarized in Figs.\,\ref{fig:velx}, \ref{fig:mix}, and Table\,\ref{tab:sn_props}. A more massive hydrogen envelope indicates a strong deceleration and reverse shock. A deeper H/He interface will result in an earlier reverse shock. Jointly these two effects will facilitate favorable conditions for hydrodynamic insabilities, resulting in more mixing and seeding clump structures. 

In Fig.\,\ref{fig:velx}, we show the projected mass-weighted velocity along the  3$\%$ $^{56}$Ni isosurface for three viewing angles (we choose positive x-, y-, and z-directions in our simulation grid) at the moment of shock breakout. We are interested in exploring the (an)isotropy of the ejecta velocity to provide insight into the shellular geometry of the explosion. The lowest mass, more symmetric 9-M$_{\odot}$ explosion shows the most isotropic ejecta velocity distribution, but even for this model the ejecta velocity is not symmetric around 0 km s$^{-1}$. Rather, the models studied here all show evidence for clumping in the $^{56}$Ni velocity distribution. The 11- and 17-M$_{\odot}$ progenitors show larger velocities and velocity distributions characteristic of their explosion asymmetry. Such clumping properties would manifest in nebular phase observations of supernova ejecta velocities, and we see that manifestation of clumping is a generic result for all the models we study here.

The top panel of Fig.\,\ref{fig:mix} shows the relation between explosion energy and $^{56}$Ni ejecta dipole fraction at the time of mapping to FLASH plotted against progenitor core compactness, $\xi_{1.75}$. Generally, higher compactness progenitors have higher explosion energies (compare with Fig.\,9 in \citealt{burrows2023}, which shows the ejecta mass dipole), mixing is affected by the interplay of explosion energy with compactness, and compactness is strongly correlated with the envelope binding energy and outer structure \citep{burrows2018,burrows_2020,burrows2024}.  Hence, not surprisingly, we can discriminate low and high-compactness models ($\xi_{1.75}$ $<$ 0.3 and $\xi_{1.75}$ $>$ 0.7) by their ejecta morphology. We conclude that within each of these compactness intervals explosion energy and dipole fraction track compactness.

Note that the 11-M$_{\odot}$ and 17-M$_{\odot}$ models represent local maxima in both explosion energy and dipole strength within each of these respective compactness neighborhoods. These models also have the highest mixing fraction, despite the former not having a particularly high explosion energy (both also have similar hydrogen envelope masses, but at different radii). We attribute this to (locally) large explosion energies, with blasts propagating through a massive envelope located deeper in the star. 

We also explore the dependence of mixing fraction on the progenitor structure. Generally, stars with a higher core compactness have a larger envelope mass beginning further out. We see a strong monotonic trend between core compactness and the H/He interface radius, both of which are critical to the development of the reverse shock, RTI, and $^{56}$Ni mixing. Despite their high compactnesses, the 23-M$_{\odot}$ and 25-M$_{\odot}$ models have a smaller envelope mass than neighboring models\footnote{by $\sim$1 M$_{\odot}$ when compared with that of the 17-M$_{\odot}$ and 20-M$_{\odot}$ models and comparable to that of the 9-M$_{\odot}$ model}, and thus have less mixed $^{56}$Ni. Model 23-M$_{\odot}$ has a lower-mass envelope and a progenitor H/He interface at larger radius, both of which result in weaker metal mixing into the hydrogen envelope. Despite a similarly located H/He interface, the 20-M$_{\odot}$ progenitor has a greater mixed fraction at breakout than the 23-M$_{\odot}$ progenitor, and we attribute this to a larger envelope mass (by $\sim$1 M$_{\odot}$).  Model 25-M$_{\odot}$ has a deeper H/He interface than model 23-M$_{\odot}$ and a slightly higher mixed $^{56}$Ni fraction. Such branching in progenitor properties with progenitor mass has previously been identified \citep{sukhbold2018,wang}. 

We also find a branched trend between explosion energy and $^{56}$Ni mixing, where more energetic explosions do not necessarily yield more efficiently mixed $^{56}$Ni. This degeneracy in explosion energy and mixed $^{56}$Ni fraction is largely resolved by the degree of asymmetry of the synthesized $^{56}$Ni. $^{56}$Ni that is more asymmetrically produced by explosive nucleosynthesis through the end of the neutrino-driven explosion is more heavily mixed outwards (top right panel of Fig.\,\ref{fig:mix}). Generally, the amount of mixing is roughly correlated with the breakout time, since the H/He interfaces expand faster into the low-density environment than the ballistic $^{56}$Ni.  




We lastly also investigated the origin of the $^{56}$Ni ejecta velocity distribution (Fig. \ref{fig:velx}). We illustrate in the bottom panels of Fig.\,\ref{fig:mix} a strong monotonic trend between both the maximum $^{56}$Ni velocity and the spread in $^{56}$Ni velocity (defined as the width of the velocity distribution encompassing 10$\%$ of the maximum velocity) with the progenitor compactness as well as the explosion energy. 
\vspace{-0.6cm}
\subsection{Explosive Burning vs. Pre-collapse Nucleosynthesis}

\begin{figure*}
\centering
    \includegraphics[width=0.41\textwidth]{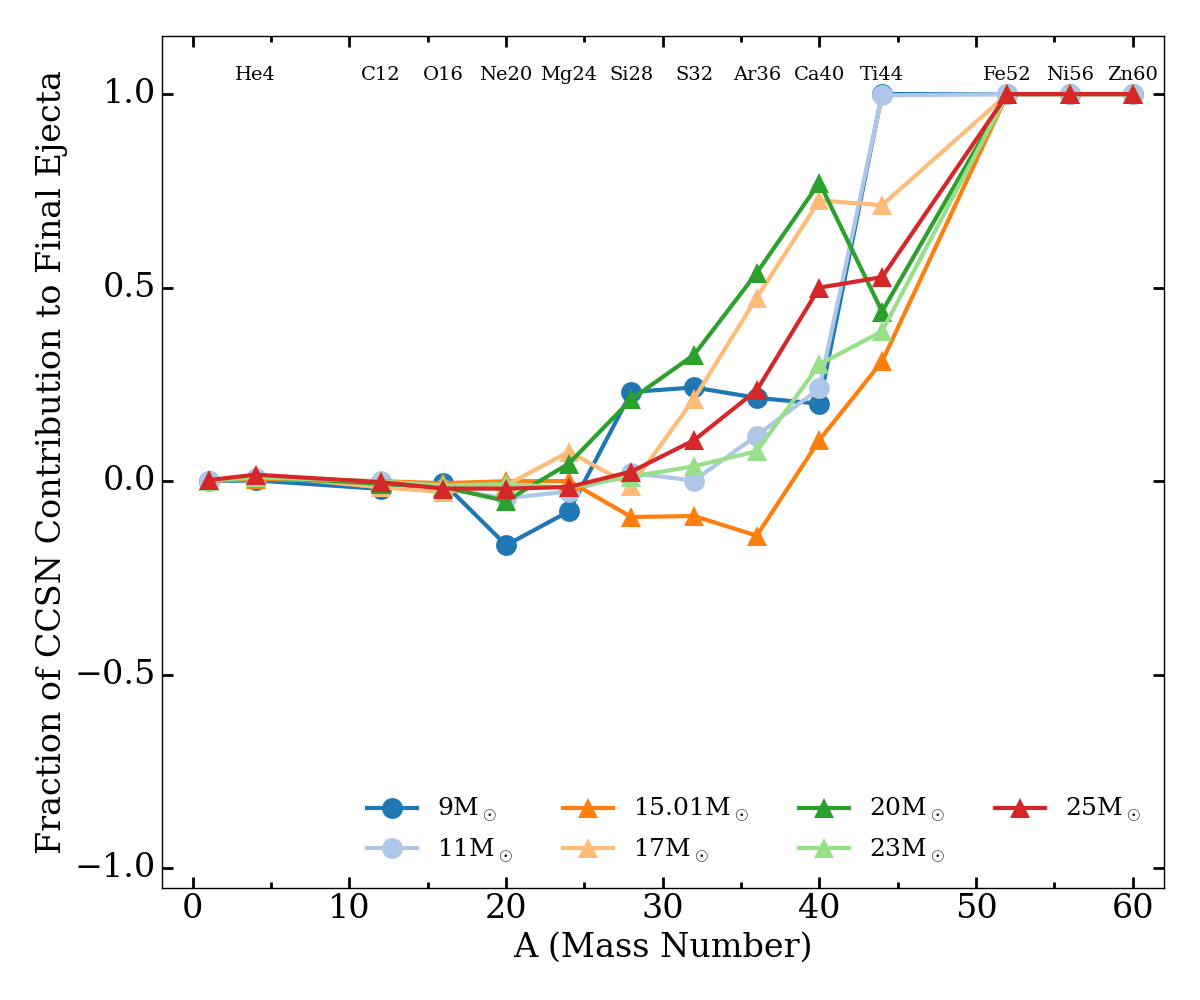}
   \includegraphics[width=0.41\textwidth]{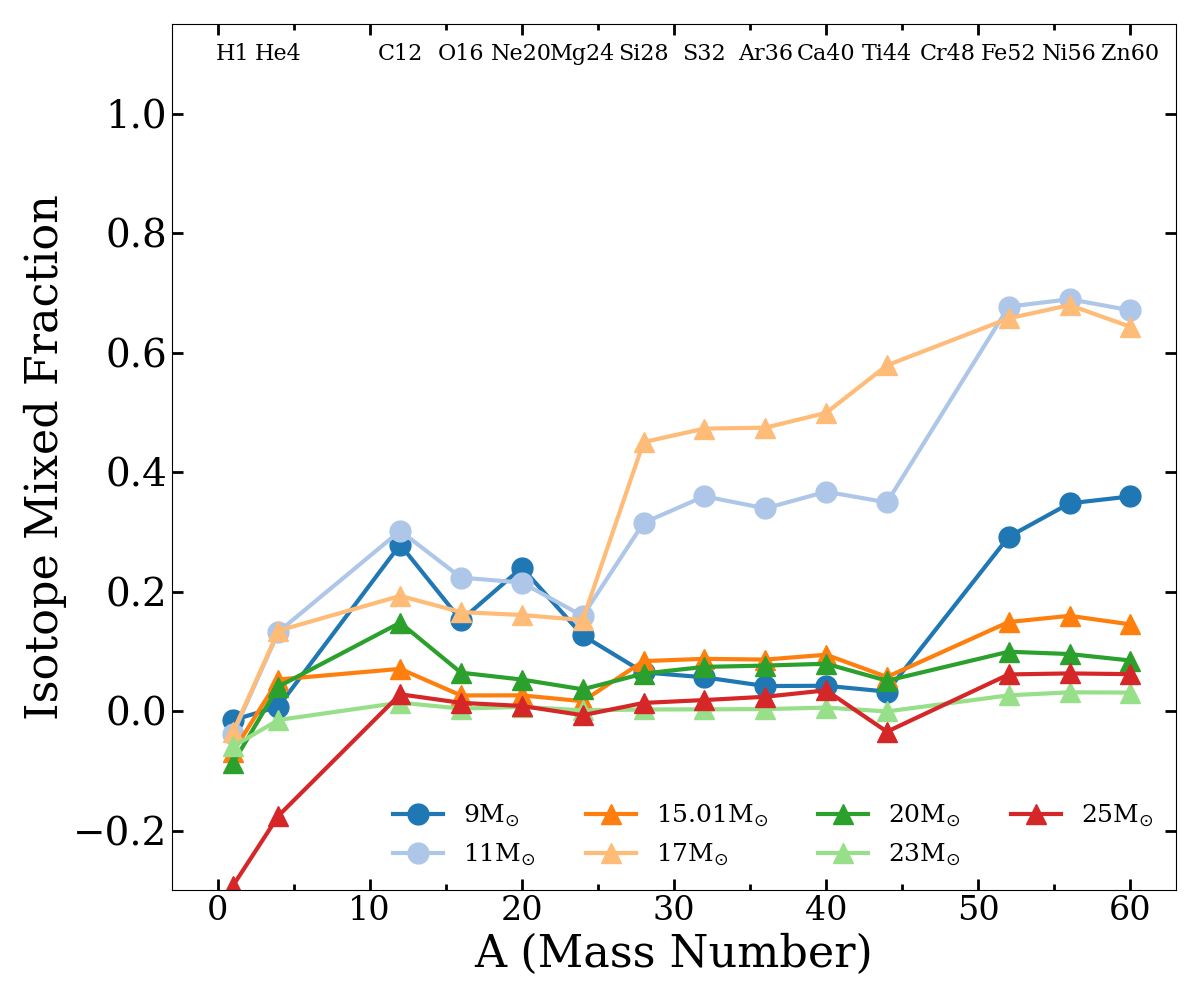}
   \caption{
   \textbf{Left:} The fraction of explosion nucleosynthesis contribution to the total yields (as opposed to pre-collapse burning) from CCSNe explosion (see also \protect{\citealt{2024ApJ...962...71W}}, Fig.\,1) by mass number A in the first seconds of the neutrino-driven explosion. \textbf{Right:} The mixed fraction of elements at shock breakout by mass number A for the seven models studied here. Generally, more massive progenitors with higher explosion energies result in greater explosive nucleosynthesis yields (beyond $^{16}$O) compared to synthesis through stellar evolution. However, elemental mixing into the outer envelope betrays a strong dependence on progenitor structure as we discuss throughout.  Note that the 9-, 11-, and 17-M$_{\odot}$ models show significant $^{56}$Ni mixing, but the 9-M$_{\odot}$ lacks significant mixing between silicon ($^{14}$Si) and $^{56}$Ni. On one hand, heavily-mixed,  low-mass/low-energy CCSNe progenitors and, on the other hand, energetic but weakly-mixed progenitors both will lack metal signatures in their gaseous envelope, with implications for nebular spectroscopy. Conversely, energetic and well-mixed explosions (the 11- and 17-M$_{\odot}$ models) show a distinct bump starting at silicon, where explosive burning begins to dominate element production. We highlight the tradeoff between explosion energy and progenitor structure.}
    \label{fig:mixed_bo}
\end{figure*}

We briefly explore the dependence of mixing of elements into the gaseous RSG envelope on both the isotope and progenitor. In Fig.\,\ref{fig:mixed_bo}, we illustrate the fractional contribution from explosive nucleosynthesis at the end of our F{\sc{ornax}} simulations to the elemental yields from pre-collapse burning. We compare this to the mixing fraction of isotopes into the base of the hydrogen envelope at the time of shock breakout. Generally, more massive progenitors with higher explosion energies result in greater explosive nucleosynthesis yields (beyond $^{20}$Ne) compared to synthesis through stellar evolution \citep{burrows2023,2024ApJ...962...71W}. However, elemental mixing into the outer envelope conveys a strong dependence on progenitor structure, as we discussed throughout. We describe key characteristics here.

We observe some mixing of lighter elements, particularly C/O, as they bleed into the H/He interface due to their proximity to the reverse shock. This is weaker for models with a later reverse shock. Next, we see a dip into elements that are neither produced explosively in significant amounts nor are near to the H/He interface. Then, we see that a clump of alpha elements above $^{24}$Mg are mixed only if they are produced mostly due to explosive burning (right panel in Fig. \ref{fig:mixed_bo}). Lastly, we see that a span of elements near the iron-peak in the 9-, 11-, and 17-M$_{\odot}$ models do show significant mixing. Note that the 9-M$_{\odot}$ model lacks significant mixing between silicon and $^{56}$Ni because of a smaller contribution from explosion nucleosynthesis of these elements. On one hand, heavily-mixed, low-mass/low-energy CCSNe progenitors and, on the other hand, energetic, but weakly-mixed progenitors, both lack enhanced metal signatures in their gaseous envelopes, with implications for nebular spectroscopy. Lastly, we find that inward mixing of H and He is more pronounced in more massive progenitors.\footnote{Elements, like $^{44}$Ti and $^{48}$Cr, are produced during the later wind phase in $\alpha$-rich freeze-out \citep{wind} and not just via explosive nucleosynthesis. Our current FLASH simulations are missing this late-time contribution.} 



    

\section{Conclusions}\label{sec:conc}

The morphology of breakout for red supergiant progenitors of Type IIp supernovae follows a general trend for all models. In the first hundreds to thousands of seconds, the asymmetries in the $^{56}$Ni structure become more accentuated as the matter moves down the density gradient. Upon first interaction with the hydrogen/helium interface (consistent with the variation in $\rho$r$^{3}$ and mass pile-up), the $^{56}$Ni ejecta become less asymmetric as the most extended features are compressed. Thereafter, RTI instabilities form, the fastest $^{56}$Ni bullets emerge through the reverse shock, and the structure of the ejecta again becomes more asymmetric. In our RSG models, the dominant mixing episodes occur at the H/He interface. 

For our sample of RSG progenitors, we find that massive stars with higher explosion energies, greater neutrino-driven asymmetries, larger hydrogen envelopes located deeper in experience greater elemental mixing into the hydrogen envelope at greater velocities. We did not find a clear correlation between any of these individual quantities or pairings and mixing yields.

We observe a tradeoff between explosion energy and H/He interface location in determining mixing and ultimate ejecta velocity. Models with a higher compactness and higher explosion energies generally boast gaseous envelopes further from the stellar core and this inhibits effective metal mixing into the envelope.  Models with a more-extended H/He interface with a smaller envelope mass mix out a smaller fraction of $^{56}$Ni mixed by shock breakout. The 11- and 17-M$_{\odot}$ models, as local maxima in an explosion energy versus compactness plot, have the highest mixing fractions at shock breakout. 

We observe a non-monotonic dependence of metal mixing into the ejecta with explosion energy and explosive burning $-$ models with higher explosions energies and greater $^{56}$Ni yields do not necessarily mix $^{56}$Ni into the base of the hydrogen envelope more efficiently. In addition, we observe differential mixing of elements; elements produced through explosive nucleosynthesis, as opposed to pre-collapse burning, are more thoroughly mixed outward. We observe that the remnant kick direction is anti-aligned with the $^{56}$Ni distribution for most of our models, with exceptions for the 9- and 23-M$_{\odot}$ progenitors.
We emphasize that, despite looking for rigorous trends, we have found only general classifications. Our study, despite being the largest sample of 3D progenitors studied through shock breakout from the stellar surface, is still limited in its sample size and we rely on a particular set of progenitor models \citep{swbj16,sukhbold2018}. The correlation between the various progenitor properties, including stellar mass, density profile, compactness, and envelope properties, is not sufficiently constrained. Determining the mapping between the first seconds of a neutrino-driven explosion and the morphologies and mixing at shock breakout is tricky and depends upon the explosion details and the aforementioned progenitor properties in a complex way. The explosion energy, dipole structure, remnant kick velocity, and compactness alone are insufficient to fully characterize the detailed ejecta structure and kinematics. 

Continuing our supernova simulations after shock breakout from the stellar surface, we find large-scale structures in the $^{56}$Ni ejecta. We do not see significant metal mixing post-breakout into a circumstellar medium, at least for the low-mass CSM environments considered here. $^{56}$Ni continues to inertially coast outward during the first day after breakout. The surrounding less dense fractions too penetrate into the CSM, but soon all mixing abates. Thus, the mixed $^{56}$Ni fraction is largely set at shock breakout for the less-dense CSM environments we have considered. We do not explore in detail the inward H and He mixing post-breakout. We are developing a more systematic follow-up study of the role of a parametrized CSM in potentially shaping explosion properties.

Three dimensional simulations such as we have presented here may provide a natural resolution of the O/Mg problem, wherein existing spherically-symmetric supernova models either overproduce oxygen or underproduce magnesium \citep{2021ApJ...921...73G}\footnote{Note the similar three-peaked profile in their Fig. 9 as found in our Fig.\,\ref{fig:mixed_bo}.}, leading to an excess in the ratio of oxygen to elements between magnesium and nickel. As we show in Fig.\,\ref{fig:mixed_bo}, differential mixing of elements may naturally addresses this issue. Oxygen, being formed almost entirely during the pre-collapse phase, with some burning during core collapse, is less outwardly mixed and will likely experience greater fallback than elements such as silicon and nickel, which are formed in greater abundance through explosive burning and, hence, are more easily mixed outward\footnote{Though this may depend upon the choice of inner boundary condition, see \citealt{2025ApJ...987..164B,wongwathanarat2015}.}.

Energetic and well-mixed explosions (the 11- and 17-M$_{\odot}$ models) show a distinct bump starting at magnesium/silicon, where explosive burning begins to dominate element production. Note also for two models (yet again the 11- and 17-M$_{\odot}$), we observed both appreciable contribution from explosion burning to the magnesium yields and a few 10$\%$s mixing into the H/He envelope (although significant mixing begins beyond silicon). Thus, explosive burning and subsequent effective mixing could enhance magnesium in the ejecta. However, neon is more effectively mixed outward than magnesium in all our models, and, thus, pre-collapse mixing may also be a necessary ingredient to explain large Mg/Ne ratios (e.g., in N49B \citealt{2025ApJ...984..185S}). 


In this paper, we have focused on RSG progenitors of Type IIp supernovae. The development of reverse shocks and compositional mixing will differ greatly for mass stripped progenitors that may yield Type Ib/c SNe \citep{2018MNRAS.479.3675M}, and we will explore this in a follow-up study. 

Spherically-symmetric, ad-hoc explosions, while certainly valuable of their own right, are simply insufficient to capture the complicated emergent structure, element mixing and dominant asymmetries. Our early simulated light curves, in development, following shock breakout for our RSG Type IIp SN progenitors show significant variation in the peak of the light curve with viewing angle, given the $\sim$day timescale variation revealed in 3D shock breakout (see also \citealt{vartanyan2025}). Additionally, metal mixing into the hydrogen envelope will be visible in nebular spectra, which should also be sensitive to varying lines-of-sight. How thoroughly and extensively the asymmetric ejecta is mixed will also effect the onset time, degree, and development of polarization in supernovae.

The possibility that distinctive signatures in the light curves and spectra during the first weeks to months in the development of a supernova's light curve and spectra could reveal distinctive signatures of heavily-mixed $^{56}$Ni, prior to the nebular phase, remains an exciting prospect. In particular, ejecta clumping would affect emission line strengths, complication progenitor mass estimations \citep{2021A&A...652A..64D}, and yield an earlier nebular phase \citep{2018A&A...619A..30D}. Regardless, the ejecta morphology, velocity, and distribution will inform current and upcoming spectropolarimetric observations to provide insight into the central engine of explosion. 

\section*{Data Availability}
We will make the angle-averaged ejecta data public and provide the reduced breakout hydrodynamic profiles, including isotope distributions, at https://dvartany.github.io/data/. More complete 3D profiles of data presented in this paper can be made available upon reasonable request to the first author. 

\section*{Acknowledgments}

DV is grateful for valuable discussions with Michael Gabler and Anthony Piro. We are also grateful to Joseph Insley (ALCF) A.B. would like to thank the Space Telescope Science Institute under grant JWST-GO-01947.011-A for its generous support. DV acknowledges support from the NASA Hubble Fellowship Program grant HST-HF2-51520. DK is supported in part by the U.S. Department of Energy, Office of Science, Office of Nuclear Physics, DE-AC02-05CH11231, DE-SC0004658, and DE-SC0024388, and by a grant from the Simons Foundation (622817DK). TW acknowledges support by the U.~S.\ Department of Energy under grant DE-SC0004658 and through Simons Foundation grant (622817DK). This work benefited from collaborations through the Gordon and Betty Moore Foundation through Grant GBMF5076  We are happy to acknowledge access to the Frontera cluster (under awards AST20020 and AST21003). This research is part of the Frontera computing project at the Texas Advanced Computing Center \citep{Stanzione2020}. Frontera is made possible by NSF award OAC-1818253. Additionally, a generous award of computer time was provided by the INCITE program, enabling this research to use resources of the Argonne Leadership Computing Facility, a DOE Office of Science User Facility supported under Contract DE-AC02-06CH11357. Finally, the authors are pleased to acknowledge that the work reported on in this paper utilized Princeton University's Research Computing resources, and our continuing allocation at the National Energy Research Scientific Computing Center (NERSC), which is supported by the Office of Science of the U.~S.\ Department of Energy under contract DE-AC03-76SF00098.

\begin{deluxetable*}{lcccccccc}[t]
\tablecaption{
Summary table of the progenitor, supernova, and shock breakout properties of this 3D model suite.}
\label{tab:sn_props}
\tablecolumns{9}
\tablewidth{0pt}
\tablehead{}
\startdata
\multicolumn{9}{c}{}\\
M$_{\rm ZAMS}$ & $\xi_{1.75}$ & Radius & E$_{\rm exp}$ &  M$_{\rm Ni}$ & t$_{\rm map}$ & t$_{\rm bo}$  & $f_{\rm mix, Ni}$ & v$_{\rm mean, Ni}$ \\
(M$_{\odot}$)  & & (10$^8$ km) & (B) & (10$^{-2}$ M$_{\odot}$) & (s, pb) & (day) & ($\%$) & (km\,s$^{-1}$) \\
9b & 6.7$\times$10$^{-5}$ & 2.86 & 0.094 & 0.61 & 1.95  &  $\sim$1.8$-$2.3 & 35 & 350 \\
11 & 0.12 & 3.96  & 0.14 & 1.04 & 4.56  &   $\sim$0.9$-$2.2 & 69 & 1700 \\
15.01 & 0.29 & 6.26  & 0.29 & 5.42  & 4.38  &  $\sim$3$-$4  & 16 & 800 \\
17 & 0.74 & 7.03  & 1.10 & 10.0 & 5.56  &  $\sim$1$-$2 & 69 & 3200   \\
20 & 0.79 & 7.94 & 0.88 & 9.94 & 6.34  &  $\sim$2.3$-$2.8  & 10 & 1500  \\
23 & 0.74 & 9.31  & 0.51 & 8.77 & 6.20 &   $\sim$3.6$-$4 & 3 & 1900   \\
25 & 0.80 & 9.92  & 1.10 & 16.80 & 6.32  &  $\sim$2.1$-$3.1 & 7 & 2400   \\ \hline
\hline
\enddata
\tablenotetext{}{\textbf{Note:} We summarize key characteristics of the progenitor models, early explosion diagnostics, and asymptotic properties of the explosion ejecta. We include the progenitor model ZAMS (zero-age main sequence) masses, their core compactness and radius, the explosion energies and nickel masses, the time of mapping to FLASH and the range and distribution of shock breakout times, and the mixed fraction of the $^{56}$Ni ejecta into the stellar envelope and their average velocities at breakout. }
\end{deluxetable*}

\clearpage
\newpage
\bibliographystyle{aasjournal}
\bibliography{References}

\label{lastpage}
\end{document}